\documentclass[sigconf]{acmart}
\AtBeginDocument{%
  }

\usepackage{enumitem}
\usepackage{multirow}
\usepackage{caption}
\usepackage{subcaption}
\usepackage[most]{tcolorbox}
\usepackage{todonotes}
\usepackage{makecell}

\newlist{questions}{enumerate}{2}
\setlist[questions,1]{label=RQ\arabic*:,ref=RQ\arabic*}
\setlist[questions,2]{label=(\alph*),ref=\thequestionsi(\alph*)}

\usepackage{xcolor}
\usepackage[normalem]{ulem}

\copyrightyear{2026}
\acmYear{2026}
\acmConference[ICER 2026 Vol. 1]{Proceedings of the ACM Conference on International Computing Education Research Vol.1}{August 11--14, 2026}{Uppsala, Sweden}
\acmBooktitle{Proceedings of the ACM Conference on International Computing Education Research Vol.1 (ICER 2026 Vol. 1), August 11--14, 2026, Uppsala, Sweden}

\renewcommand\footnotetextcopyrightpermission[1]{}

\begin{document}

\newtcolorbox{finding}{
  enhanced,
  sharp corners,
  colback=white,             
  colframe=gray,            
  boxrule=0.5pt,
  left=3pt,
  right=3pt,
  top=3pt,
  bottom=3pt,
  before skip=3pt,
  after skip=3pt,
  fonttitle=\bfseries\small,
  coltitle=black,
  attach title to upper={},
  separator sign={\ },
}

\title{Characterization and Effects of CS2 Learning with GenAI, Visualization, and Human Support}

\author{Quinton Yong}
\email{quintonyong@uvic.ca}
\affiliation{%
  \institution{University of Victoria}
  \department{Computer Science}
  \city{Victoria} 
  \state{British Columbia}
  \country{Canada}
}

\author{Anthony Estey}
\email{aestey@uvic.ca}
\affiliation{%
  \institution{University of Victoria}
  \department{Computer Science}
  \city{Victoria} 
  \state{British Columbia}
  \country{Canada}
}

\author{Miguel Nacenta}
\email{nacenta@uvic.ca}
\affiliation{%
  \institution{University of Victoria}
  \department{Computer Science}
  \city{Victoria} 
  \state{British Columbia}
  \country{Canada}
}

\begin{abstract}
    Generative AI (GenAI) is becoming a widely adopted learning support tool for both students and instructors, as it offers benefits such as personalized tutoring and scaffolded learning. However, recent research highlights potential drawbacks such as overreliance and metacognitive issues, especially in novice programmers. Most prior work focuses on introductory programming courses, and important questions remain about the underlying mechanisms behind the negative effects of GenAI and if findings can be generalized when students learn more advanced computer science concepts. To address this gap, we conducted a mixed-methods study comparing student interactions with GenAI to two traditional learning supports in a second-year algorithms course: algorithm visualization (AV) and human live tutoring (LT). Twelve students participated in three 90-minute study sessions focusing on sorting, tree, and graph algorithms. We recorded gaze and interaction data, and each session concluded with a test assessing their conceptual understanding of the topic. Our analysis classifies when during the problem-solving process participants sought help, and compares the interaction patterns across the three learning supports. Although GenAI produced a larger increase in self-efficacy compared to live tutoring, it was associated with noticeably lower results in learning outcomes. We found that participants did not use algorithm visualizations effectively, faced usage barriers when using GenAI to learn advanced topics, and that live tutoring yielded the highest learning outcomes.

\end{abstract}

\ccsdesc[500]{Social and professional topics~Computer science education}
\ccsdesc[300]{Human-centered computing~Human computer interaction (HCI)}

\keywords{CS2, generative AI, LLMs, algorithm visualization, human tutors, self-efficacy, metacognition}


\maketitle

\section{Introduction}

The emergence and adoption of Generative AI (GenAI) tools is rapidly transforming computer science (CS) education~\cite{CSedinEraofGenAI}. The increasing capabilities of GenAI tools (e.g.,~\cite{CodexCS2}) creates opportunities for increased support of learning \cite{CS50_1} but also introduces pedagogical challenges \cite{chatgptThreats}. This changing technological landscape is forcing students and instructors to adapt rapidly, sometimes without much knowledge of what does and does not work.
We are still unsure of how GenAI affects learning~\cite{wideninggap} and the consequences of its integration into classrooms~\cite{AITEachingAssitant}. Research has shown that GenAI tools have the potential for scaffolding learning~\cite{genAIscaffolding}, making learning support more accessible~\cite{CodeAid, IrisAITutor}, and increasing self-efficacy~\cite{studentAIinteraction}. However, research has also raised concerns regarding academic integrity~\cite{academicintegritygenai}, overreliance~\cite{overreliance1}, and exacerbation of the metacognitive awareness issues faced by first-year computer science students~\cite{wideninggap}. 

Of particular concern is the lack of knowledge about GenAI supports in more advanced algorithms and data structure courses (CS2). Most existing studies focus on introductory CS and programming courses (e.g.,~\cite{focusonintroductoryprogramming}), whereas only a few address CS2 \cite{fewercs2, GenAIinupperyear} and advanced algorithms courses \cite{rev_genAIalgscourse}. Yet, CS2 is difficult~\cite{cs2difficulties, cs2dropoutrate} and may represent a critical point affecting student retention in computer science programs \cite{rev_CS2inflection1, rev_CS2inflection2, rev_CS2inflection3}. Therefore, it is important to know if the challenges encountered when adopting GenAI to support CS1 courses also generalize to CS2.

In this paper we consider whether results from existing literature extend to CS2 by performing a mixed-methods study of twelve students learning topics in algorithms and data structures. 
Our study design facilitates fine-grained analysis of student interactions with learning supports through screen recordings and eye-tracking data.
To better understand the mechanisms of GenAI's effect on learning these topics, we compare student interactions with GenAI to those with a human tutor and to interactions with another traditional form of learning support, namely algorithm visualization~\cite{AVstateofthefield}. We are unaware of such previous comparisons.
We incorporate an analysis of student learning outcomes, what students perceive as the most effective learning supports, and also explore how these align with student self-efficacy and metacognition.

Our main contributions are the results derived from our analysis. We found that self-efficacy increased most after students interacted with the algorithm visualization tool, somewhat with the GenAI support, and least with the live tutor. However, the resulting measures of learning were highest after live tutoring and lowest after using GenAI. Additionally, learners preferred live tutoring to GenAI, and ranked algorithm visualization last. Algorithm visualization was sparsely used interactively and was not used effectively. 

These results corroborate the emerging patterns regarding how GenAI affects CS education and expand them beyond CS1 contexts. Additionally, they suggest reasons why algorithm visualization's potential is still relatively unfulfilled, and may offer insights to inform the design of next-generation learning support systems.

\section{Related Work}
\label{sec:relatedwork} 
We consider three groups of relevant previous literature: how GenAI is being used in CS education, how GenAI affects self-efficacy and metacognition, and how visualization has been used to support understanding of algorithms.

\subsection{Generative AI as a Learning Support in Computer Science Education}
The growing popularity and widespread adoption of Generative AI (GenAI) tools is transforming Computing Science education~\cite{robotsarehere, CSedinEraofGenAI}. GenAI can, with minimal or no human intervention, solve exam questions and exercises from both introductory programming (CS1)~\cite{robotsarecomingintroprogramming, GPT4IntroCS} and more advanced algorithms and data structures courses (CS2)~\cite{CodexCS2}. 

GenAI can support CS learning by providing explanations of code or creating programming exercises~\cite{ComparingCodeExplanations, ProgrammingExercises}. Researchers have integrated GenAI into programming classes \cite{AITeachingSystematicReview}, while implementing pedagogical guardrails to promote learning \cite{CS50_1, CS50_2, LLMHelpRequest, CodeHelp}, and investigated the effects of GenAI integration on learning outcomes in CS1 \cite{rev_genAICS1learningoutcomes}. GenAI-based CS education tools are motivated as a scalable support for programming~\cite{CodeAid, CodeHelp}, for automated grading \cite{rev_AIgrading}, or as teaching assistants \cite{AITEachingAssitant}. GenAI-based tutor systems \cite{CS50_1, IrisAITutor} have also been developed to provide accessible \cite{rev_genAItutoraccessible} and personalized instruction \cite{rev_genAItutordialogue}. However, researchers have also described concerns about the negative effects of GenAI on education, including how it intensifies academic dishonesty challenges~\cite{academicintegrity, academicintegritycs50, academicintegritygenai}, reduces problem solving or critical thinking skills~\cite{ProgrammingUsedToBeHard, llmcriticalthinking}, and might result in overreliance of AI tools~\cite{overreliance1, overreliance2, rev_genAICS1learningoutcomes}.

To facilitate the pedagogically effective incorporation of GenAI into classrooms and reduce its potentially negative effects, another stream of research aims to understand student interactions with GenAI. Prather et al. \cite{interactionchallengesprather} and Nguyen et al.~\cite{interactionchallengesnguyen} investigated interaction challenges between novice programmers and GenAI tools which observed issues of struggling to generate effective prompts, being confused by outputs, and misunderstanding GenAI's capabilities.
Qualitative analyses of GenAI interactions such as Adeeb et al.'s \cite{interactionscodetracing} and Kazemitabaar et al.'s \cite{interactionspythoncourse} found that beginner programmers using GenAI use problem-solving strategies ranging from completely writing the solution themselves to fully relying on AI output, and observed that students showed signs of both self-regulation and overreliance.

Despite the many interesting findings in existing work, we believe there are still some significant gaps.
Larger scale studies which incorporate AI into classrooms can provide a large body of student survey or course outcome data, but can only overview how students individually use the tools in practice. Smaller scale interaction studies provide fine-grained analysis, but do not look at learning outcomes directly influenced by the GenAI. Studies which aim to use GenAI as tutors have not compared these systems with actual human tutors. We aim to fill these gaps by conducting fine-grained comparative analysis on the interactions of students with GenAI and human tutors, while also measuring direct learning outcomes.

\subsection{The Effect of GenAI on Self-Efficacy and Metacognition}

Bandura defined self-efficacy as an individual's belief in their ability to achieve specific goals~\cite{selfefficacydefinition}. Empirical evidence supports that self-efficacy is a strong predictor of success in CS education~\cite{selfefficacypredictor1, selfefficacypredictor2}. Instruments exist to measure the self-efficacy of learners~\cite{MSQL} and also specifically of novice programmers~\cite{CS1selfefficacysurvey}. Danielsiek et al.~\cite{algselfefficacy} developed the A-SES (Algorithms Self-Efficacy Scale) which is an instrument to assess self-efficacy specifically for students in algorithms courses, and found that the main factors influencing self-efficacy were related to algorithm design, advanced paradigms, runtime analysis, and pseudocode writing/tracing. Recent studies examine how GenAI tools affect the self-efficacy of students. Margulieux et al.~\cite{selfregulationselfefficacy} observed that students in introductory programming courses who used AI more frequently and earlier in programming tasks had lower self-efficacy than students who used AI less frequently and later in tasks. Amoozadeh et al.~\cite{studentAIinteraction} studied CS1 students' interactions with GenAI when solving programming tasks and observed that some students' self-efficacy increased after using GenAI, but that completely relying on GenAI correlated with lower self-efficacy. 

Another aspect of cognition which has implications for CS education is metacognition~\cite{metacognitiondefinition}, an individual's ability to reflect and regulate their own cognitive process. Metacognition is likely an important aspect in learning CS~\cite{metacognitioninprogramming}. Prather et al.~\cite{metacognitivedifficulties1} investigated the metacognitive awareness challenges which CS1 students face (e.g. forming incorrect mental representation of a problem and keeping incorrect partial solutions).
A follow-up study~\cite{wideninggap} showed that GenAI could exacerbate existing difficulties and introduce new metacognitive difficulties (e.g. being misled by the tool). The study also found that students with lower self-efficacy experience more metacognitive difficulties.

Existing studies of GenAI which also look at self-efficacy do not include measures on student learning outcomes and have thus far focused only on novice programmers. We aim to broaden our understanding by considering the three-way relationship between GenAI, self-efficacy, and outcomes, and by examining whether these challenges also extend to CS2.

\subsection{Algorithm Visualization as a Support Tool}
Algorithm Visualization (AV) tools have been historically proposed as a method to facilitate learning and teaching of algorithms and data structures~\cite{surveyofsuccessfulAV, oldAV} through visual representations and animations of how algorithms work~\cite{AVstateofthefield}. Many have proposed AV tools for learning the mechanisms of algorithms~\cite{visualgohalim, AP-ASD1, jhave, AVstateofthefield, vizalgo, opendsa}. Other visualization tools, namely program visualization systems~\cite{rev_PVsurvey, programvisualization}, support the learning of programming and debugging through code execution visualization  (e.g.,~\cite{pythontutor, jeliot, ville, pinti}) and can also be applied to algorithm learning~\cite{rev_PVinalgscourse}. Studies on visualization tools show that visualizations can facilitate learning~\cite{AVmetastudy}. Naps et al.~\cite{AVengagement} suggest that the effectiveness of visualizations depends on the level at which learners engage with them (\textit{viewing}, \textit{responding}, \textit{changing}, \textit{constructing}, \textit{presenting}), with more active forms of engagement being more effective for learning. AV can also reduce cognitive load~\cite{avcognitiveload1, avcognitiveload2} and increase motivation and self-efficacy~\cite{AVselfefficacy}. However, there are still questions regarding the pedagogical effectiveness of AVs \cite{AVmetastudy} and an ongoing unanswered problem is the lack of adoption of visualization tools despite its claimed benefits~\cite{AVnotused1, AVnotused2}. By comparing student interactions with AV to interactions with GenAI, which is a widely adopted learning tool, we contribute to understanding the effective mechanisms of AV tools \cite{AVeffectiveness, AAVeffectiveness} and also inform how GenAI's pedagogical effectiveness can be improved.

\section{Research Goals and Questions}\label{goalsandRQs}
Although Section~\ref{sec:relatedwork} shows that there is clear progress in our understanding of how new GenAI technologies affect and can be integrated into the learning of computer science, we identified gaps that we believe are important for our understanding and development of future versions of learning support technologies. We therefore directed our research towards the following specific research goals (RG):
RG-A) We aim to generalize previous findings on GenAI support beyond first-year computer programming courses, as these courses are different because they focus more on algorithm understanding and analysis than on the basic primitives of programming and implementation (i.e., we aim to complement and extend research which focuses exclusively on first-year programming courses~\cite{robotsarecomingintroprogramming, wideninggap, studentAIinteraction}); RG-B) We aim to provide comparisons with natural baselines of learning supports, anchoring the results on existing (pre-GenAI) knowledge (i.e., we focus on a comparative analysis rather than considering only GenAI or newly proposed systems in contrast to work such as~\cite{interactionchallengesprather,interactionscodetracing}),
and; RG-C) We aim to offer a fine-grained view of individual learners behavior and connect them to learning outcomes directly affected by GenAI so that we can extract possible solutions to problems and suggest plausible causes to issues (in contrast to learner self-reported measurements as in~\cite{CS50_1, IrisAITutor, CodeAid} or aggregate grade measurements as in~\cite{wideninggap}).
These research goals are instantiated as four main research questions:
\begin{questions}[start=0]
    \item How do learners interact with different learning supports when trying to understand CS2 concepts?\label{RQ:0}
    \item How does learner self-efficacy change after using different learning supports?\label{RQ:1}
    \item Do learning supports affect learners' understanding of the CS2 concepts differently?\label{RQ:2}
    \item What do learners perceive as the most effective learning support?\label{RQ:3}
\end{questions}

\section{Empirical Study}
We designed a semi-controlled empirical study to answer the research goals and questions from the previous section. The design of our study is specifically tailored to the goals and research questions above; we link the design to their corresponding research goal below when appropriate. The research questions and the main analysis procedures were pre-registered to improve replicability and because we support avoidance of HARKing~\cite{harking}. The pre-registration is accessible on OSF.\footnote{\url{https://osf.io/vxfrg/overview}. We note that H4 became H0 in the paper for clarity.} The study was pre-approved by the local human research ethics review board.

\subsection{Participants}
We recruited twelve volunteer students from a group of approximately 125 students 
enrolled in a data structures and algorithms course at the University of Victoria (RG-A). 
Participants responded to an in-person class announcement and a follow-up message in the class learning management system. Participants received \$45 payment.
None of the authors were involved in the teaching or marking of the course. The training that participants underwent as part of the study was relevant for the learning goals of the course. The integration of an actual course with the study was aimed at increasing the ecological validity of the results.

\subsection{Study Design Apparatus and Procedure}
The experiment followed a within-subjects design in which each participant experienced each of the three learning supports once in three separate training sessions to enable cross-supports comparison (RG-B). The main factor is the learning support: Live Tutor (LT), Algorithm Visualization (AV) and a generative artificial intelligence chat interface (GenAI). The training sessions were modeled after the tutorials or laboratory sessions that are common in this kind of course in 
computer science departments, but they took place one-on-one with the experimenter (live tutor) in an instrumented private office with a PC set up with an Eyelink 1000 Plus\footnote{\url{https://www.sr-research.com/eyelink-1000-plus/}} desktop-mounted gaze tracker which allows free head movement. 

Each session covered one CS2 topic aligned with the learning outcomes of the course, but not directly covered in the course content. The three topics were \textit{Counting Sort}, \textit{AVL Trees}, and \textit{Unweighted Graph Shortest Paths}, always in that order. The sessions were scheduled after students covered sufficient relevant material in the course for each topic (i.e., after the week on search algorithms, binary search trees, and graph representation respectively). The sessions did not require full understanding of course material to complete the study tasks. The order in which participants experimented each condition was counterbalanced (two participants in each of the six possible orders). Out of the 36 sessions, 35 sessions were completed (P10 could not attend their corresponding LT session).

\begin{figure}[h] 
    \centering 
    \includegraphics[width=\linewidth]{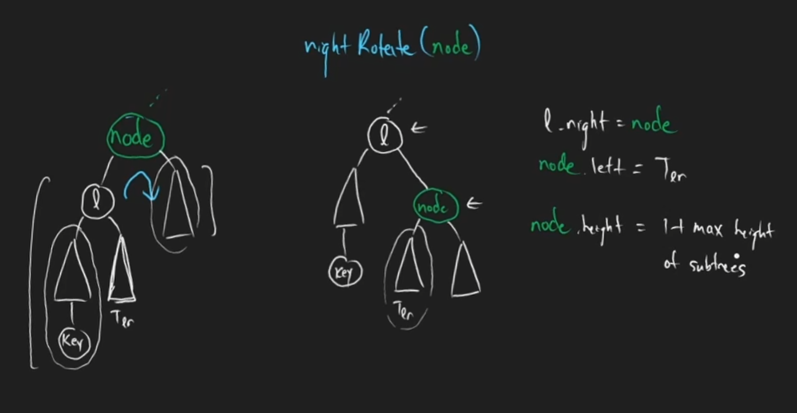}
    \caption{ A frame of the Session 2 video on AVL trees.} 
    \label{fig:video} 
\end{figure}

The sessions lasted approximately 90 minutes each. Each session started with a description of the session process, then continued with the following phases: A) a pre-session self-efficacy questionnaire; B) the calibration of the gaze-tracker; C) a lecture video on the session topic (e.g., an explanation of AVL trees---see Figure~\ref{fig:video}); D) a post-video questionnaire; E) an exercise to complete with the corresponding learning support; F) a post-exercise questionnaire; G) a quiz to assess understanding; H) a post-test self-efficacy survey, and; I) a survey where the student reported on the learning support's usefulness. The first session had an additional demographic survey and signing of consent, and the last session added a student preference ranking survey of all learning supports.  
The full instruction materials, the quizzes, and all survey instruments are available as part of the supplementary materials attached to the pre-registration.\footnote{\url{https://osf.io/q6ac7/overview}}

\begin{figure*}[h] 
    \centering 
    \includegraphics[width=1\textwidth]{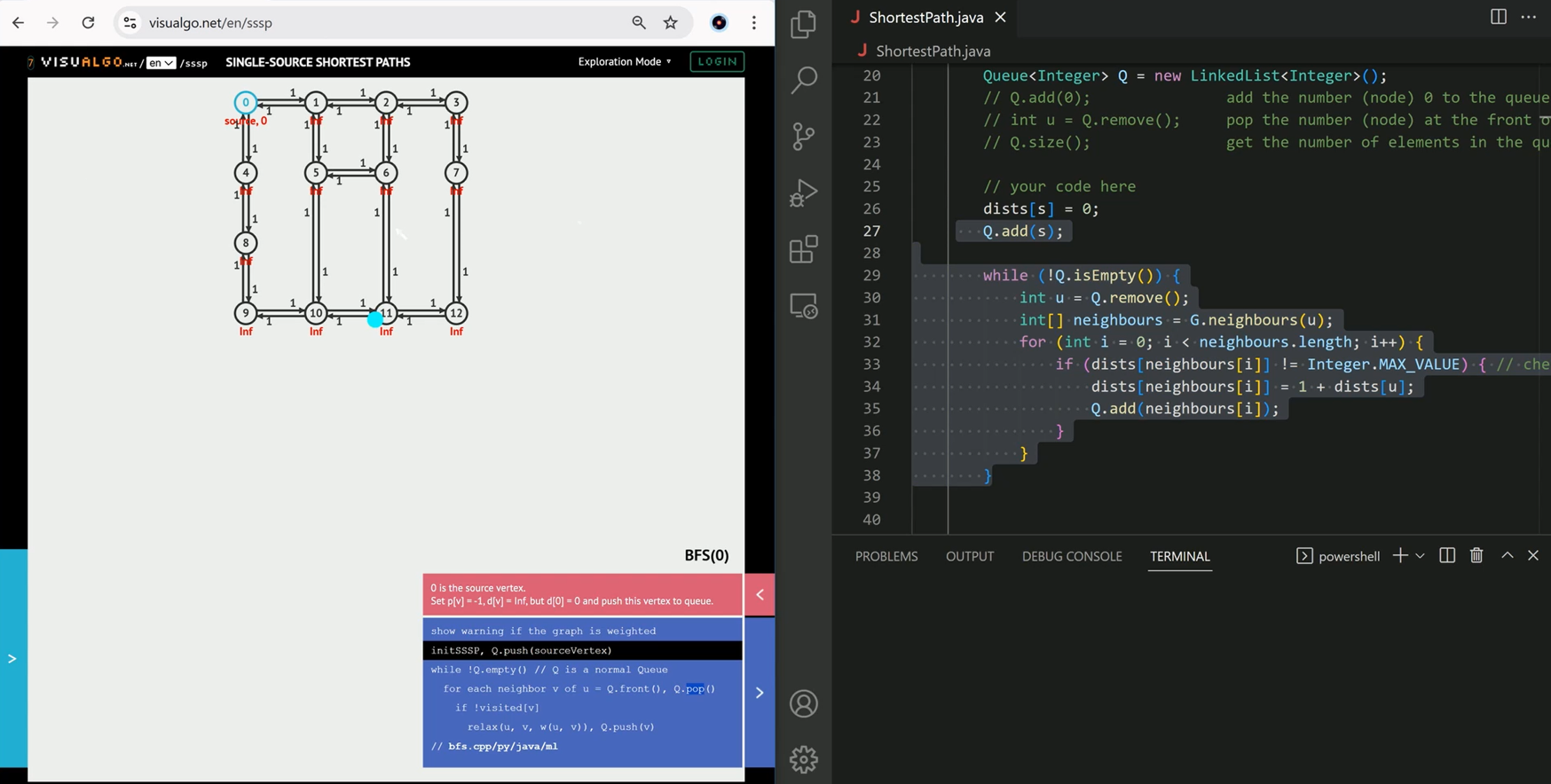}
    \caption{Desktop capture of the Session 3 exercise on unweighted graph shortest paths with the AV tool support.} 
    \label{fig:desktop} 
\end{figure*}

The core of each session were the lecture (phase C---between 15 and 25 minutes), where participants watched a pre-recorded video in which the first author explained the algorithm or data structure with the help of a virtual digital blackboard, and the subsequent exercise training (phase E---between 10 and 50 minutes), where learners tried to understand the topic and implement it with the help of a learning support. To complete the exercise, the PC showed the VSCode editor set up with a Java compiler (the same language used for programming assessments in the corresponding course) with a function skeleton of the assigned topic and a testing framework. Participants were also given very high level pseudocode descriptions (serving as notes from the lecture video) in an adjacent window as well as the assigned learning support (in the case of GenAI or AV). A snapshot of the desktop setup is illustrated in Figure~\ref{fig:desktop}. During the exercise, participants were allowed to refer back to the lecture video and to use pen and paper.

\subsection{Conditions}

The main difference between conditions was the availability of a different learning support. Besides the timely study of GenAI support, we selected algorithm visualization as a prominent example of computer supported learning that has been extensively studied (see Section~\ref{sec:relatedwork}), and live human tutoring, which still represents the status quo in personalized instruction and is a suitable baseline for comparison (RG-B). Each of the conditions were presented in the following manner:

\paragraph{Live Tutor (LT)} In this condition, participants were instructed to ask for assistance from the first author in the same way that they would in the lab sessions of the course. The first author has been a tutor several times for this course and other similar courses. 

\paragraph{Algorithm Visualization (AV)} The PC included a window with the appropriate algorithm or data structure visualization in the  VisuAlgo~\cite{visualgohalim} AV tool. We selected VisuAlgo (as opposed to, e.g.~\cite{pythontutor, pinti}) because it is a sophisticated system which represents the main functionalities of state-of-the-art AV tools, it is programming language-agnostic, it includes the algorithm and data structure topics which are covered in most CS2 courses, and it is easily accessible online to students. 
VisuAlgo enables learners to set up parameters and conditions for the algorithm and execute a particular problem step-by-step.

\paragraph{Generative AI (GenAI)} Instead of the VisuAlgo window, participants saw an instance of Microsoft's Copilot large language model (LLM) chat interface (at the time running a GPT-4–class model).
Participants could cut and paste text or images between any of the windows, as they would on their own PCs. We selected Copilot because it was the most powerful LLM available through our institution at the time and because we believe it is representative of the GenAI tools that our students currently use for learning support in self-guided study. We acknowledge that there are more sophisticated agents specifically geared towards tutoring (e.g.~\cite{CS50_1, CodeAid, IrisAITutor}, although not all are easily accessible), yet we consider Copilot a reasonable baseline system. 

\subsection{Measurements}
To achieve a detailed look at learner behavior (RG-C), we planned the collection of a wide range of measurements of outcome and behavior data. 

First, we collected \textbf{objective learning measurements} through the quiz in phase G of each session. The quiz is the main measure of learning outcomes, the construction of which was informed by Thompson et al.'s work on applying Bloom's taxonomy to computer science assessment \cite{bloomstaxonomycs, bloomstaxonomy}. Each session had a specific quiz that tested knowledge on the topic of the session, but the quiz taken was identical for the three learning support conditions. The quizzes contained five questions specifically designed to test the learners' ability to: 1) trace the algorithm or data structure operations; 2) modify the algorithm for a problem variation; 3) analyze the running time; 4) describe edge cases; and 5) compare the algorithm or data structure to one they have previously learned. Each question was graded out of 2, where 2 corresponded to a fully correct answer, 1 if the provided solution had a minor error, and 0 if the solution had at least one major error. Participants could not use the lecture video or the learning support for answering the quiz. 

To connect with previous research, we also assessed \textbf{self-efficacy} before and after the learning core of the session (in phases A and H). Self-efficacy measurements were a modified version the Algorithms Self-Efficacy Scale (A-SES) questionnaire~\cite{algselfefficacy} with 5-Likert scale questions.
The second self-efficacy questionnaire used the same questions, but the scale was extended to 7 options so that learners who had selected the maximum value in either direction on the previous scale could express further movement in either direction after learning. The main objective of this measurement is to assess the change in self-efficacy as an effect of the learning session (as per the pre-registration).

Another post-session questionnaire used 5-level Likert questions (informed by the factors influencing self-efficacy found by Danielsiek et al. \cite{algselfefficacy}) which explicitly asked learners about the \textbf{utility of the learning support} for different aspects of learning algorithms. Specifically,  1) Designing an algorithm 2) Understanding how an algorithm or data structure works 3) Analyzing running time 4) Implementation 5) Tracing execution.

We collected a continuous \textbf{screen video capture with audio} of phases C and E that we automatically annotate with \textbf{gaze location} to enhance our understanding of which material or learning support tool participants were using at a given time. Because the gaze tracking device cannot record when participants are looking out of the screen, the experimenter also recorded \textbf{instances of use of pen and paper}. The audio allows us to capture verbal interactions with the live tutor/experimenter, as well as any participant comments during GenAI and AV condition sessions. A first level of analysis of the video allows us to produce \textbf{counts of interactions} with the editor, the video lecture, and the learning supports, as well as measurements of \textbf{time spent on each}. The survey at the end of the last session, when participants had experienced all the conditions, allowed us to capture \textbf{comparative preferences} between the learning supports. 

We also collected two additional surveys (phases D and F) but these were primarily used to validate aspects of the experiment and the manipulations and were not central to the analysis, hence we will not report on them.
The materials used for all measurements as well as the numerical data collected through all the surveys above are in the supplementary materials (attached to the pre-registration).

\subsection{Analysis}
In Section~\ref{sec:Results} below, we report summary statistics and analysis for the quantitative measurements and the details of the thematic analysis approach for the qualitative analysis. 

\subsubsection{Quantitative}
By default, quantitative test results are associated to hypotheses that were pre-registered before data collection. We label tests or data manipulations that were not pre-registered. 

For quantitative hypothesis testing we use an MCMC Bayesian framework based mostly on Kruschke's approach~\cite{rjags,kruschke_bayesian_2021} because: a) It circumvents or ameliorates conceptual and practical problems with traditional null hypothesis statistical testing (NHST---see, e.g.,~\cite{kline_beyond_2013,cumming_new_2014}); b) It offers more transparency regarding the statistical models used, and; c) It supports a more nuanced interpretation of test probabilities. Our analyses are implemented in JAGS 4.3.1~\cite{plummer2003jags}
on R 4.4.3~\cite{Rlanguage}. Every tested parameter was visually tested for good mixture of the Markov chains and achieved at least ESS of 10,000 samples ($0.995 < \text{psrfs} < 1.005$). 
We also performed post-predictive visual tests of the models. To facilitate interpretation, we consider hypotheses or comparisons for which the probability, given the data, is above 95\% to be \emph{strongly supported} and between 90\% and 95\% to be \textit{weakly supported}. For correlations, we estimate the correlation slope and consider that there is a correlation if the 95\% High-Density Probability Interval of the value does not include zero. We report High-Density Probability Intervals (HDI) for each parameter of interest. HDIs are credible intervals containing 95\% of the probability mass (see~\cite{rjags}) which contain the mode of the parameter if distributions are unimodal (ours are).

The model for condition comparisons is based on a Gaussian model of error (Equation~\ref{eqn:error}), with the average model as a linear function of the condition and the participant (Equation~\ref{eqn:anova}):

\begin{equation}\label{eqn:error}
P(y \mid c,p) = \frac{1}{{\sigma\sqrt {2\pi } }}e^{{{ - \left( {y - \mu(c,p) } \right)^2 } \mathord{\left/ {\vphantom {{ - \left( {y - \mu(c,p) } \right)^2 } {2\sigma^2 }}} \right. \kern-\nulldelimiterspace} {2\sigma^2 }}}
\end{equation}

\begin{equation}\label{eqn:anova}
\mu(c,p) = \beta_0 + \beta_1(c) + \beta_2(p)
\end{equation}
where $y$ is the outcome variable of interest (e.g., test score), $\sigma$ is the estimated standard deviation of the error, $c$ is the main condition (GenAI, AV or LT), and $p$ is the participant. The hypotheses are comparisons of the plausible estimations of parameter $\beta_1$, such as calculating the probability, given the data that $\beta_1(LT) > \beta_1(AV)$.

The estimations of correlation are based on a Student-t distribution of error (more robust than the normal distribution):
\begin{equation}\label{eqn:errorcorr}
P(y \mid x,c) = StudentT(\mu(x,c),\sigma,\nu)
\end{equation}

\begin{equation}\label{eqn:anovacorr}
\mu(x,c) = \beta_0 + \beta_1(c)\cdot x
\end{equation}
where $x$ and $y$ are the correlated measurements, $c$ is the condition, $\sigma$ is the estimated standard deviation of the error, and $\nu$ is the degrees of freedom for the Student-t distribution.

Priors for all relevant parameters are chosen to reflect our lack of previous information and are, therefore, highly uninformative and easily overwhelmed by the data.
The parameters in the condition comparison models ($\beta_0$, the three $\beta_1$, and the 16 $\beta_2$) have normal prior distributions centered on zero and with standard deviations five times the standard deviation observed in the data overall. The prior for $\sigma$ is a uniform distribution between 0.01 and 1000. 
For the correlation models, the priors of $\beta_0$ and the three $\beta_1$ are normal distributions centered on zero and with standard deviations ten times the standard deviation of the corresponding dataset. The prior for $\sigma$ is also a uniform distribution between 0.01 and 1000, and for $\nu$ is a broad exponential with rate parameter $\lambda = 30$.

We provide all the data, the JAGS model specifications, and the R scripts necessary to run them in the supplementary materials.

\subsubsection{Qualitative}
The coding and qualitative analyses were performed on the screen recording data (which included overlaid gaze-tracking data) and audio recordings. The codebook was developed using both deductive and inductive approaches. The initial codebook  included 11 categories. Five categories were informed by the difficulties of CS2 summarized Mtaho et al.~\cite{cs2difficulties} and included \textit{support-specific difficulties} for each of the three learning supports, \textit{misunderstanding of algorithm / data structure concept}, and \textit{difficulty with solution implementation}. Three categories were informed by the programming problem solving stages by Loksa et al. \cite{barriersinproblemsolving} and included \textit{evaluation before implementation}, \textit{solution implementation}, and \textit{evaluation of implemented solution}. Another three categories were \textit{queries directed to the three respective learning supports}. Initial codes in each category were populated based on specific participant behaviours observed by the first author during the sessions.

Coding was done iteratively. The first author initially coded the recordings from the first two sessions of each condition, and tagged each learning support interaction, problem-solving stages, and observed difficulties (facilitated by eye-tracking data). Following each coding pass, all researchers met to verify code applications and discuss and resolve ambiguities in code interpretation. During these discussions, additional inductive codes were introduced to capture emergent support interaction behaviours, difficulties with learning supports and implementation, and shifts in attention between learning supports. Previously coded data were then revisited to incorporate the new codes. This iterative process was repeated three times until all sessions were coded and a final version of the codebook was established (included in the supplementary materials), after which the first author performed an additional coding pass through all the session data using the finalized codebook. Following the coding phase, the authors collaboratively grouped codes into higher level themes \cite{thematicanalysis} to identify recurring interaction patterns, problem solving strategies, and difficulties related to the exercise and learning support usage.

\section{Results}
\label{sec:Results} 

We organize the results according to the research questions in Section~\ref{goalsandRQs}. First, we characterize the general patterns of use with each of the three learning supports. Then, we examine changes in self-efficacy, student performance, correlations, and the participant's own reports on the effectiveness of the learning supports.

\subsection{Interactions with Learning Supports}

We examined in detail each interaction that each participant had with each of the three learning supports in phase~E
of the study. We classified each interaction based on a high-level categorization of the interaction's purpose that we inferred iteratively from the recordings and gaze behavior. Although we considered using categorizations from earlier work (e.g.,~\cite{interactionchallengesprather, wideninggap, interactionspythoncourse}),
we found that the categories focused heavily on beginner programmer related topics or  challenges, were too difficult to infer from the data, or were only partially relevant for the current task.
We arrived at our own categorization based on six categories for the live tutor and the GenAI and three categories for the algorithm visualization. The derived categories for the live tutor and GenAI are described, with examples, in Table \ref{tab:categoriesltgenai}. 
The algorithm visualization support tool required a separate set of categories because of intrinsic differences in how students can interact with it. Specifically, it is much harder to infer from interactions with the AV, which are mostly non-verbal, whether a participant is querying algorithm understanding or finding an error. Other categories are also non-applicable; for example the AV is intrinsically unable to assist with syntax issues. 
Descriptions and examples of each of the these categories are shown in Table \ref{tab:categoriesav}. 

\begin{table*}[h]
\renewcommand{\arraystretch}{1.2}
  \centering
\caption{Names, descriptions, and examples for each of the categories of queries to the live tutor and GenAI.}
\label{tab:categoriesltgenai}
\begin{tabular}{ |p{1.8cm}|c|p{5.7cm}|p{5.6cm}| } 

\hline  
 \multicolumn{2}{|c|}{\textbf{Query Category}} & \multicolumn{1}{c|}{\textbf{Description}} & \multicolumn{1}{c|}{\textbf{Example}} \\
\hline
\textit{Algorithm Understanding} & \textit{AU} & Interactions querying about the fundamental mechanisms of the algorithm & P5 asked the LT, ``In the algorithm, is there a reason why we traverse the array backwards instead of forwards?''\\ 
\hline
\textit{Code Writing} & \textit{CW} & Interactions querying specifics on how to write code snippets, subroutines, or portions of the algorithm implementation &  P3 asked the GenAI, ``Can you show how to calculate the height [of a tree] in Java?'' \\ 
\hline
\textit{Verification of Correctness} & \textit{VC} & Interactions to verify if a particular statement (usually corresponding to their understanding of how a part of the algorithm or its implementation works) is correct &  P2 asked the LT, ``Is this [way I'm accessing this node] valid for temporarily saving it?'' \\ 
\hline
\textit{Syntax} & \textit{SN} & Any specific queries about code syntax & P10 asked the GenAI, ``How do I return an empty array in Java?''  \\ 
\hline
\textit{Error Finding} & \textit{EF} & Queries that specifically refer to locating or fixing a bug or error in the code &  Upon encountering failed tests cases, P6 copied their function into the GenAI with no additional query.  \\ 
\hline
\textit{Problem Clarification} & \textit{PC} & Meta-queries about the exercise itself & P12 asked the LT, ``Do we have to consider cases where the keys are the same?''  \\ 
\hline
\end{tabular}
\end{table*} 

\begin{table*}[h]
\renewcommand{\arraystretch}{1.2}
  \centering
\caption{Names, descriptions, and examples for each of the categories of queries to the AV tool.}
\label{tab:categoriesav}
\begin{tabular}{ |p{1.9cm}|c|p{5.6cm}|p{5.6cm}| } 

\hline  
 \multicolumn{2}{|c|}{\textbf{Query Category}} & \multicolumn{1}{c|}{\textbf{Description}} & \multicolumn{1}{c|}{\textbf{Example}} \\
\hline
\textit{Step Tracing} & \textit{ST} & Participants watch a sequence of animated steps corresponding to a part of the algorithm & P4 stepped through the initial stage of a sorting algorithm to see how an auxiliary array is populated.\\ 
\hline
\textit{Test Case Creation} & \textit{TC} & Participants create a new test as input to the algorithm to visualize the its execution &  P5 created a test case for the section of the AVL tree function they were implementing \\ 
\hline
\textit{Statically Visualize} & \textit{SV} & Participants spend a non-trivial amount of time looking at a static image of the data structure in the AV tool & P8 looked at a still image of the graph in the AV tool for an extended period while trying to find their code error \\ 
\hline

\end{tabular}
\end{table*}

\subsubsection{Volume and Types of Interactions}

The full counts of all interactions is shown in Table~\ref{tab:interactions}. The distribution of interaction types offers an initial coarse picture of how interactivity patterns differ between learning supports. Although interactions with AV are not comparable in type, we attempted to count interactions in a consistent way across techniques so that the numbers of interactions would be broadly comparable. This was only possible because we collected eye-tracking data. For example, a group of different subsequent clicks through the AV would only count as one interaction, and this interaction continued until the participant shifted their focus (e.g., moved to updating their code).
Equivalently, a back and forth interaction on the same topic with the GenAI or the live tutor would only be counted as one interaction, until the purpose of the interaction changed, the participant solved the issue, or the issue was abandoned altogether.

\begin{table*}[h]
\renewcommand{\arraystretch}{1.3}
  \centering
\caption{Interaction counts for the three learning supports, according to the types in Tables~\ref{tab:categoriesltgenai} and~\ref{tab:categoriesav}. The ``I'' columns represent aggregates of interactions.}
    \begin{tabular}{l|cccccc|c|cccccc|c|ccc|c|}
    
    \cline{2-19}         
    & \multicolumn{7}{c|}{\textbf{LT}} & \multicolumn{7}{c|}{\textbf{GenAI}} & \multicolumn{4}{c|}{\textbf{AV}}  \\
    \cline{2-19}
    \textbf{} & \multicolumn{1}{c}{\textbf{AU}} & \multicolumn{1}{c}{\textbf{CW}} & \multicolumn{1}{c}{\textbf{VC}} & \multicolumn{1}{c}{\textbf{SN}} & \multicolumn{1}{c}{\textbf{EF}} & \multicolumn{1}{c|}{\textbf{PC}} & \multicolumn{1}{c|}{\textbf{I}} & \multicolumn{1}{c}{\textbf{AU}} & \multicolumn{1}{c}{\textbf{CW}} & \multicolumn{1}{c}{\textbf{VC}} & \multicolumn{1}{c}{\textbf{SN}} & \multicolumn{1}{c}{\textbf{EF}} & \multicolumn{1}{c|}{\textbf{PC}} & \multicolumn{1}{c|}{\textbf{I}} & \multicolumn{1}{c}{\textbf{ST}} & \multicolumn{1}{c}{\textbf{TC}} & \multicolumn{1}{c|}{\textbf{SV}} & \multicolumn{1}{c|}{\textbf{I}} \\
    \hline
    \multicolumn{1}{|c|}{P1} & 2     & 2     & 1     & 1     & 1     & 0     & 7   & 0     & 4     & 0     & 1     & 0     & 0     & 5     & 8     & 0     & 4     & 12   \\
    \hline
    \multicolumn{1}{|c|}{P2} & 1     & 2     & 1     & 2     & 0     & 0     & 6     & 0     & 0     & 0     & 1     & 0     & 0     & 1      & 1     & 0     & 2     & 3     \\
    \hline
    \multicolumn{1}{|c|}{P3} & 2     & 4     & 4     & 1     & 3     & 0     & 14    & 2     & 1     & 0     & 0     & 0     & 0     & 3      & 1     & 0     & 0     & 1     \\
    \hline
    \multicolumn{1}{|c|}{P4} & 0     & 0     & 0     & 1     & 0     & 0     & 1     & 0     & 0     & 0     & 0     & 0     & 0     & 0      & 2     & 0     & 0     & 2     \\
    \hline
    \multicolumn{1}{|c|}{P5}  & 1     & 3     & 6     & 2     & 1     & 1     & 14    & 0     & 0     & 0     & 1     & 1     & 0     & 2      & 1     & 3     & 0     & 4     \\
    \hline
    \multicolumn{1}{|c|}{P6} & 0     & 0     & 0     & 3     & 1     & 1     & 5     & 0     & 0     & 0     & 1     & 1     & 0     & 2      & 2     & 3     & 1     & 6     \\
    \hline
    \multicolumn{1}{|c|}{P7} & 0     & 0     & 0     & 0     & 0     & 0     & 0     & 0     & 0     & 0     & 0     & 0     & 0     & 0     & 0     & 0     & 0     & 0 \\
    \hline
    \multicolumn{1}{|c|}{P8} & 0     & 0     & 0     & 0     & 0     & 0     & 0     & 2     & 3     & 0     & 0     & 3     & 0     & 8      & 0     & 0     & 1     & 1     \\
    \hline
    \multicolumn{1}{|c|}{P9} & 1     & 1     & 0     & 3     & 0     & 0     & 5     & 0     & 0     & 0     & 3     & 0     & 0     & 3      & 0     & 0     & 0     & 0     \\
    \hline
    \multicolumn{1}{|c|}{P10} &   -    &    -   &   -    &   -    &    -   &   -    &   -    & 0     & 0     & 0     & 2     & 0     & 0     & 2     & 0     & 0     & 0     & 0     \\
    \hline
    \multicolumn{1}{|c|}{P11} & 0     & 0     & 0     & 0     & 0     & 0     & 0     & 0     & 0     & 0     & 2     & 0     & 0     & 2     & 3     & 0     & 3     & 6     \\
    \hline
    \multicolumn{1}{|c|}{P12} & 2     & 1     & 0     & 1     & 0     & 1     & 5     & 0     & 0     & 0     & 0     & 0     & 0     & 0     & 0     & 0     & 2     & 2     \\
    \Xhline{3\arrayrulewidth}
    \multicolumn{1}{|c|}{\textbf{Sum}}  & 9     & 13     & 12     & 14     & 6     & 3     & \textbf{57}     & 4     & 8     & 0     & 11     & 5     & 0     & \textbf{28}     & 18     & 6     & 13     & \textbf{37}     \\
    \hline
    \end{tabular}%
  \label{tab:interactions}%
\end{table*}

Throughout the sessions, there were more LT interactions (57 total) than AV tool interactions (37 total) and GenAI interactions (28 total). The differences are somewhat indicative of an increased willingness to interact with the human tutor (8/11 participants did so), but the relatively small number of participants means that we have to consider this difference as an indication and not solid evidence.

The most common type of queries to the live tutor were syntax, code writing, and verification of correctness queries. The pattern for GenAI is similar, with the notable exception that none of the participants queried the GenAI for verification of correctness (VC) or problem clarification (PC). 

Interactions with the AV learning support were fairly sparse, and 13 out of the total 37 interactions used the tool merely as a static representation.

\subsubsection{Interactions Timing and Patterns}
We also recorded when each participant used the learning tools. Interactions are classified as \emph{before implementation} (after watching video but before any code writing), \emph{during implementation} (during the process of completing the first version of the code), and \emph{during debugging} (after running the code at least once and encountering compiler errors or bugs). 
The total number of interactions (with associated participants) that occurred in each phase per learning support are shown in Table \ref{tab:interactionphases}.

\begin{table*}[h]
\renewcommand{\arraystretch}{1.4}
  \centering
\caption{Interaction counts and participants who performed the interaction for the three learning supports, by phase.}
\label{tab:interactionphases}
\begin{tabular}{ c|l|l|l| } 
\cline{2-4}  
 & \textbf{Before Implementation} & \textbf{During Implementation} & \textbf{During Debugging}\\
\hline
\multicolumn{1}{|c|}{\textbf{LT}} &  1 (P3) & 42 (P1, P2, P3, P4, P5, P6, P9, P12) & 14 (P1, P3, P5, P9) \\ 
\hline
\multicolumn{1}{|c|}{\textbf{GenAI}} & 0 & 18 (P1, P3, P5, P6, P8, P9, P10, P11) & 10 (P1, P2, P5, P6, P8, P9, P10) \\ 
\hline
\multicolumn{1}{|c|}{\textbf{AV}} & 0 & 26 (P1, P2, P3, P5, P6, P11) & 10 (P1, P4, P8, P11, P12) \\ 
\hline
\end{tabular}
\end{table*}

The most salient insight from Table~\ref{tab:interactionphases} is that with only one exception (P3---with LT) and for all supports, participants directly jumped into coding without queries about the algorithm or their understanding of the task. 
We interpret that most participants assumed that they knew enough to try to implement the algorithm, that the supports would not help them much if they queried, or that trying to code the algorithm was the best way to check or correct their understanding. 
There is also a noticeable discrepancy between the behaviors with GenAI and the other two supports; with GenAI the volume of interactions seems to be more balanced between the implementation and debugging stages (36\% of interactions happened in debugging) whereas interactions during debugging were more sparse with the live tutor (25\% of all interactions), and by fewer participants (only 4).

\subsubsection{Struggling with Supports}

We observed that LT interactions were fairly low-barrier, at least compared to GenAI. Most participants had LT interactions upon encountering some kind of difficulty with their implementation such as not being able to translate an algorithm step into code or being unable to identify an error. These questions were also often informally formulated; for example P5 asking how an auxiliary array is used by asking "Is that the in between one?". We suspect that participants anticipate that querying the GenAI will be harder. This is consistent with the already described behavior of delaying the question until debugging, where it would probably be easier to make the query specific. In contrast, the LT is not only (perceived to be) able to interpret more contextual or vague queries, but also to redirect participants' questions that might not be relevant to the step where the participant was stuck. For example, when implementing an AVL tree function, P12 asked how to traverse back to the root of the tree, and the LT explained how it could be done but stated that it was an unnecessary step.
 
Although querying syntax with the GenAI was straightforward and always successful, many of the other types of interactions (8 of 18 queries) were unsuccessful, with the participants not being able to progress further. The GenAI outputs were often a combination of code and additional explanations, but participants would always look first at the code and then typically ignore the additional explanations. There were three participants who used the GenAI for algorithm understanding and code writing queries, namely P1, P3, and P8. Both P1 and P3 did not make visible progress in the implementation after performing any of these queries. Notably, after P3's three queries during implementation, they stated "This is going to take me forever because it is hard for me to understand." and did not use the GenAI further. P8 interacted with the GenAI the most out of all participants in an iterative manner. However, after not being able to find their error, P8 asked the GenAI to write an entire function, along with saying "I'm doing no learning [in this session], I apologize."

The interaction data with AV suggest that this support was also difficult for participants to use effectively. In the few cases when participants went to the AV tool after encountering an issue, their experience was interpreted to be likely unsatisfactory because they often moved to the pseudocode or to rewatch the lecture video rather than editing the code, which would have signaled improved understanding or an idea of how to solve the problem. Out of the 24 step trace and test case creation interactions to the AV, 10 of these instances showed participants not progressing further. In more extreme cases of difficulties with AV, one participant during implementation was not performing the step trace in the correct algorithm step corresponding to their implementation. Another participant was repeatedly unable to create a test case which ran the part of the algorithm they were implementing. P7 noted that "[The AV tool] makes [the algorithm] seem more complicated than it actually is." While trying to use the AV tool to debug, P8 noted that "I don't even need [the AV tool] because I already understand [the algorithm], it's just telling me what I already know." Furthermore, four participants during the AV condition created their own visualizations with pen and paper, despite having access to a vetted and interactive visualization. Interestingly, these pen and paper visualizations were also useful in the other conditions, sometimes to identify their errors or to ask questions to the live tutor (see Table \ref{tab:createdvis}). 

\begin{table*}[htbp]
\renewcommand{\arraystretch}{1.3}   
  \centering
  \caption{Participants who created their own pen and paper graphical representations during the sessions in the different conditions.}
    \begin{tabular}{c|c|c|c|c|c|c|c|c|c|c|c|c|c|}
    \cline{2-14}
          & \textbf{P1}    & \textbf{P2}    & \textbf{P3}    & \textbf{P4}    & \textbf{P5}    & \textbf{P6}    & \textbf{P7}    & \textbf{P8}    & \textbf{P9}    & \textbf{P10}   & \textbf{P11}   & \textbf{P12}   & \multicolumn{1}{l|}{\textbf{Total Yes}} \\
    \hline
    \multicolumn{1}{|c|}{\textbf{LT}}    & Yes   & No    & Yes   & No    & Yes   & No    & Yes   & Yes   & Yes   &   -    & Yes   & Yes   & \textbf{8} \\
    \hline
     \multicolumn{1}{|c|}{\textbf{GenAI}} & Yes   & No    & Yes   & No    & Yes   & No    & Yes   & Yes   & No    & Yes   & No    & Yes   & \textbf{7} \\
    \hline
     \multicolumn{1}{|c|}{\textbf{AV}}    & No    & No    & No    & No    & Yes   & No    & No    & Yes   & No    & Yes   & No    & Yes   & \textbf{4} \\
    \hline
    \end{tabular}
  \label{tab:createdvis}%
\end{table*}%

\subsection{Change in Self-Efficacy}\label{sec:results:SE}

Figure~\ref{fig:diffse_violin} displays the self-efficacy change before and after using each learning support. All sessions except four resulted in participants reporting an increase in self-efficacy after completing the exercise. The AV condition resulted in the largest average increase in self-efficacy (10.5 average increase on a 133 total points scale),
followed by GenAI (8.8 average increase), and the LT condition reported the lowest increase in self-efficacy (6.5 average increase). 

The post hoc pre-registered comparisons between learning supports shows fairly strong evidence that the increase in self-efficacy for AV is larger than for the LT, given the observed data (94.7\%, which approaches our convention of strong evidence). The other two comparisons are inconclusive (the probability of GenAI larger than LT is 81.8\% and AV larger than GenAI is 77.6\%). Because some participants did not directly interact with the learning supports, we ran a parallel analysis which excluded sessions without learning support interactions. We note that this analysis is exploratory and was not included in the pre-registration. For these results, the probability of AV being higher than LT was also 94.7\%, the probability of GenAI being higher than LT was 90.2\% (weakly supported), and the probability of AV being higher than GenAI is inconclusive.

\subsection{Differences in Understanding}
\begin{figure}
\begin{subfigure}[b]{0.478\textwidth}
        \includegraphics[width=\linewidth]{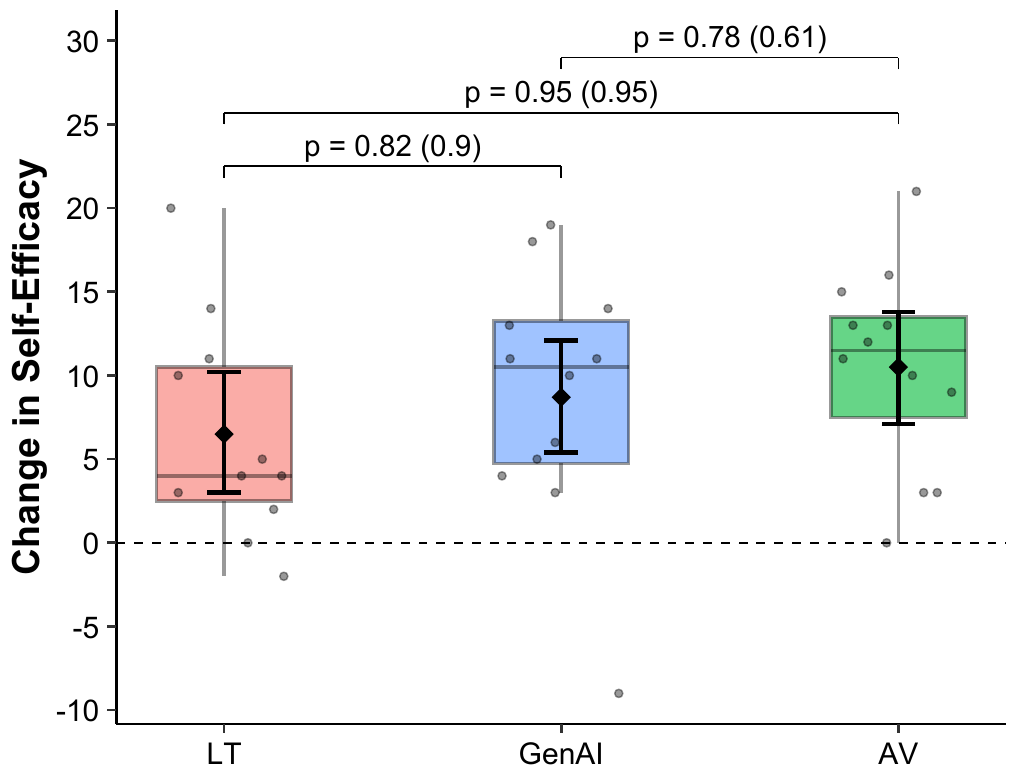}
        \caption{Change in self-efficacy by condition.} 
        \label{fig:diffse_violin} 
\end{subfigure}

\vspace{0.4cm}
\begin{subfigure}[b]{0.478\textwidth}
        \includegraphics[width=\linewidth]{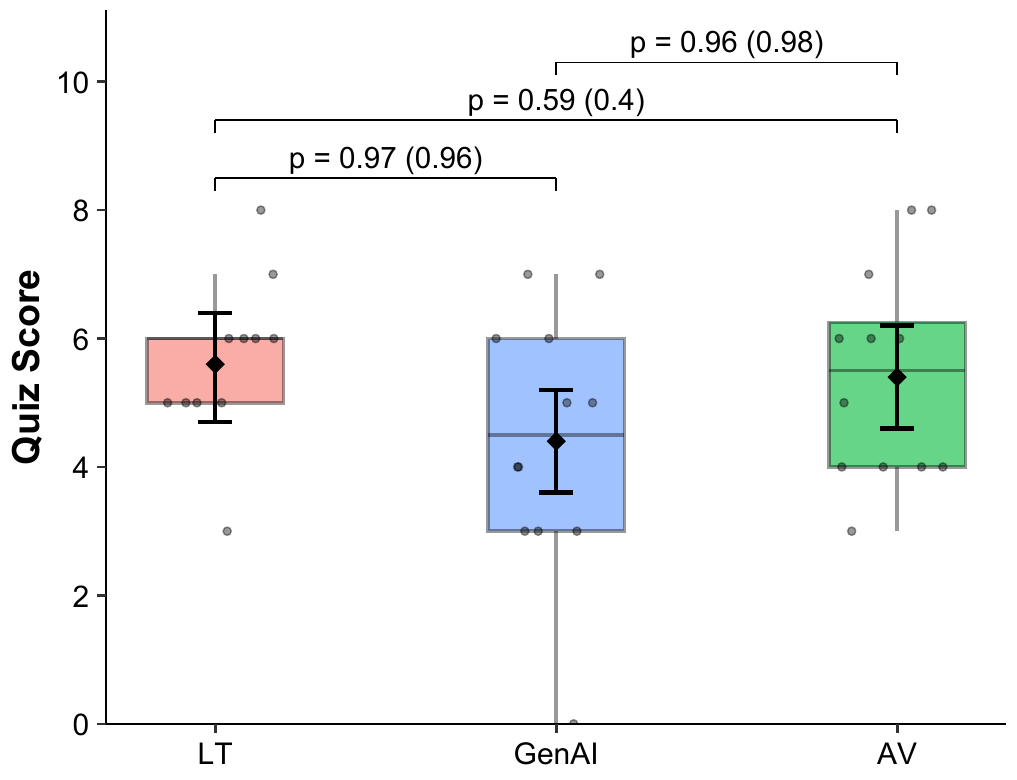}
        \caption{Quiz scores by condition.} 
        \label{fig:score_violin} 
 \end{subfigure}
        \caption{Changes in self-efficacy and quiz scores by learning support and associated probability of evidence (p values correspond to pre-registered analyses, the exploratory analyses' p are between parentheses). Boxplots in grey show medians and interquartile ranges with overlaid participant datapoints. Central diamonds represent the average and the whiskers are 95\% Probability High-Density Intervals.}
        \label{fig:SEandQuizScores}
\end{figure}

Figure \ref{fig:score_violin} shows quiz scores for each condition. The LT condition resulted in the highest quiz scores (5.6/10 average), followed by AV (5.4/10 average), with GenAI at the bottom (4.4/10 average). Post hoc comparisons show that LT scores are, on average, larger than GenAI with a probability of 97.3\%. AV scores are higher than GenAI's with $p = 96.1\%$. The difference between LT and AV is inconclusive (59.4\%). As in Section~\ref{sec:results:SE}, we ran an exploratory analysis excluding participants without learning support interactions and the results do not change the result pattern.

\subsection{Correlations}
We pre-registered linear correlation analyses to understand which measurements are related. These are run independently for each condition. Notice that none of these measurements were controlled experimentally, hence we cannot infer causality. 
We ran the following correlations:
\begin{enumerate}
    \item Change in self-efficacy vs.\ quiz scores
    \item Number of interactions vs.\ change in self-efficacy
    \item Duration of interactions vs.\ change in self-efficacy
    \item Number of interactions vs.\ quiz scores
    \item Duration of interactions vs.\ quiz scores
\end{enumerate}
In this section we report only the results for which the pre-registered tests detected a statistically reliable correlation (i.e., the estimation range of slope in the model with 95\% probability was above or below zero). A complete list of estimations is provided in the Appendix as well as the supplementary materials attached to the pre-registration. In AV, there is a positive correlation between change in self-efficacy and quiz scores (slope between [0.015, 0.409]---larger changes of self-efficacy correspond with higher scores). In GenAI, the number of interactions correlated negatively with the change in self-efficacy (more interactions also mean less self-efficacy---[-4.090, -0.860]), which also appears when we remove sessions without interaction (not pre-registered). Also in GenAI the total time spent interacting with the support correlates negatively with self-efficacy ([-0.071, -0.001]).

\subsection{Learning Tool Preferences}
\begin{figure}
     \centering
     \begin{subfigure}[b]{0.478\textwidth}
         \centering
         \includegraphics[width=\textwidth]{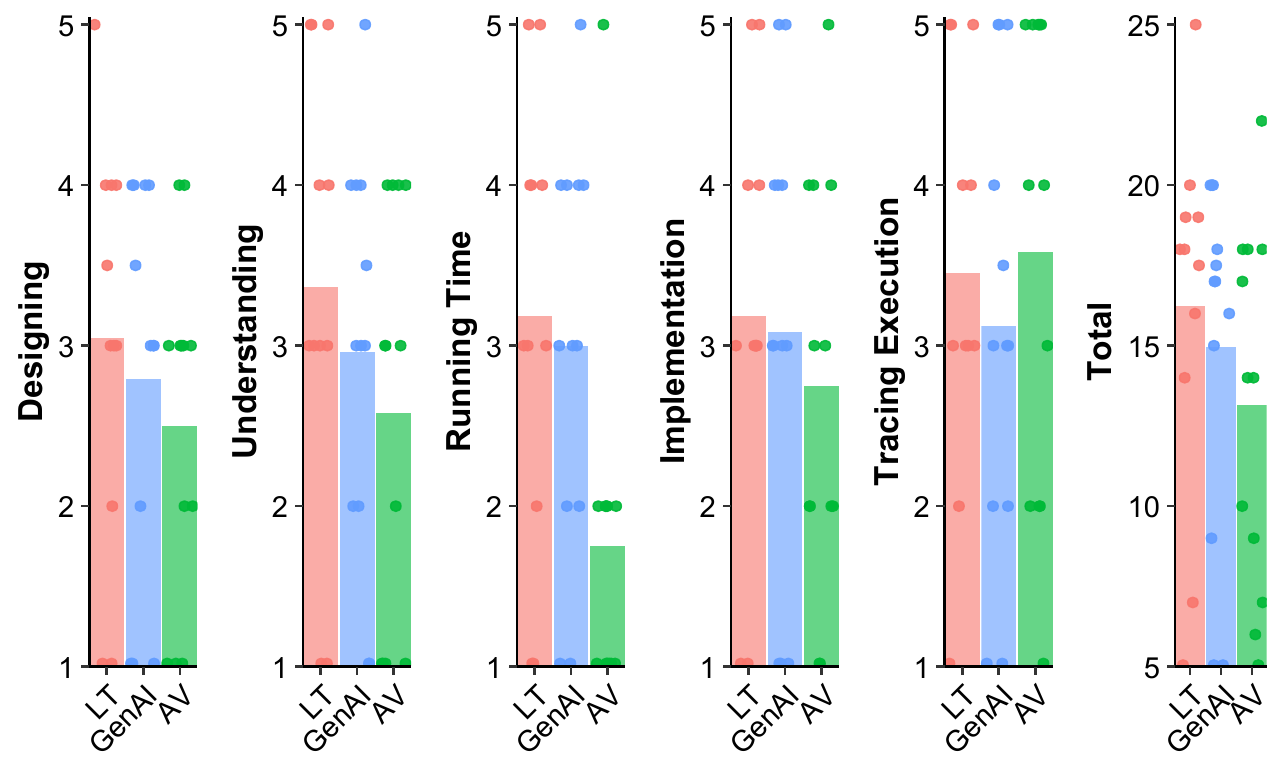}
         \caption{Learning support subjective usefulness ratings after each session.}
         \label{fig:toolusefulness}
     \end{subfigure}

    \vspace{0.4cm}
     
     \begin{subfigure}[b]{0.478\textwidth}
         \centering
         \includegraphics[width=\textwidth]{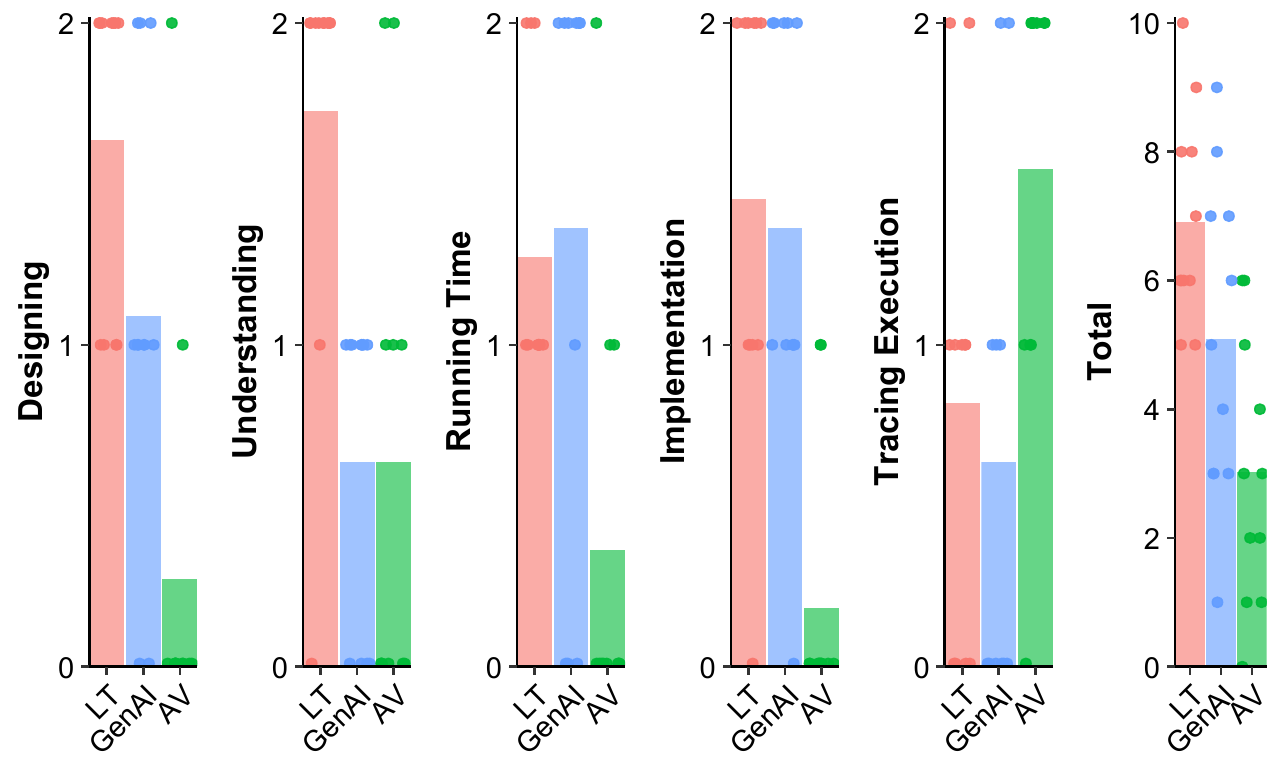}
         \caption{Rankings of subjective usefulness for learning supports after the final session. Most preferred is encoded as 2, middle as 1, and least preferred as 0.}
         \label{fig:toolfinalranking}
     \end{subfigure}
        \caption{Learning support subjective results.}
        \label{fig:toolpreferences}
\end{figure}

We collected two types of subjective participant assessments of the learning supports. First, participants rated on a 5-level Likert scale the perceived usefulness of the learning support for five different aspects of the process: 1) Designing an algorithm 2) Understanding how an algorithm or data structure works 3) Analyzing running time 4) Implementation 5) Tracing execution. This took place in phase~I
of each session and the results are displayed in Figure~\ref{fig:toolusefulness}, with the last chart showing the aggregate. At the end of the experiment, when participants had experienced all the learning supports, they also ranked each in terms of preference, using the same aspects (Figure~\ref{fig:toolfinalranking}). 

The aggregate usefulness places LT as most preferred, followed by GenAI and then AV. The pattern is similar for all aspects, except in tracing execution, where AV was rated highest, and similarly to LT. Only a few statistical comparisons in the per session usefulness scores are conclusive, which is expected because the independent subjective ratings, days apart, are understandably noisy. For a more powerful comparison we look at the final comparative rankings. These show a very similar pattern: LT is clearly rated the highest for Designing, Understanding, and overall; LT and GenAI are rated similarly for Running Time and Implementation; and AV is only rated best for Tracing Execution. The Bayesian pairwise comparisons show strong evidence for the differences highlighted above and for the aggregate comparison (the exact probabilities are in the Appendix and supplementary materials).

\section{Discussion}
In this section we summarize and interpret the results. We start with how our results respond to the research questions of Section~\ref{goalsandRQs} and follow by reviewing what we have learned about the differences between GenAI and Live Tutors, the potential issues with Algorithm Visualizations usage, and how the results fit with previous literature. 

\subsection{Answers to Research Questions}

\subsubsection{RQ0: How do learners interact with different learning supports?}
In our exercise sessions we found that, with our conceptualization of interaction, participants tended to interact more with the live tutor than with the other two supports. Although the experiment is not designed to provide definitive evidence of this, our observations suggest that barriers to initiate interactions to problem solve with a human are lower than with GenAI. This is related to Nguyen et al.'s finding that formulating helpful prompts or queries is difficult \cite{interactionchallengesnguyen} and their additional findings of tasks being cognitive demanding could be exaggerated by the increase in complexity of CS2 topics. It also seems like participants might not see it as feasible to query the AI for verification of correctness or problem clarifications, but GenAI was very effective to address syntax issues.

Participants tended to jump into implementation immediately, without revising or using the learning supports to verify their understanding,
regardless of the learning support that was offered. With the live tutor participants seemed to ``front load''  more the interactions during implementation, which meant that GenAI showed a higher proportion of queries during the later stage of debugging. 

We found that the number of participant interactions with AV was lower than LT and higher than GenAI, and the most common interactions involved more passive forms of engagement with the visualization (\emph{step tracing} and \emph{static visualization}). We observed several instances in which AV interactions did not help learners progress further through the exercise, and in some cases participants preferred to create their own pen and paper visual representations. However, step tracing being the most common type of learning support interaction, AV resulting in higher quiz scores than GenAI, and student preference of AV support for tracing the execution of algorithms suggest that AV features provide value and these features can support understanding of aspects of algorithm behaviour.
Overall, these patterns are based on empirical observations, but need to be confirmed for generality and representativeness in further studies with larger numbers of participants.

\subsubsection{RQ1: How does learner self-efficacy change after using different learning supports?}
As expected, student self-efficacy increases after hands on experience with the algorithm. However, our analysis provides some evidence that the increase is larger for AV and GenAI than with the LT. 
We speculate that interacting with another human might have been more clearly perceived as an offloading of their own learning process, compared to the other tools. We discuss this further in Section~\ref{subsec:genaivsLT} below. 
The behavior with AV that we observed was more consistent with unassisted self-learning, which might be why participants increased their self-efficacy most with this learning support (i.e., they might recognize that they did more of the work of learning, also because the support was less useful and had less scaffolding).
This hypothesis is also consistent with the negative correlation between the number and time of interactions with the GenAI and self-efficacy increase. More interactions with GenAI meant less confidence increase, perhaps because they perceived those interactions less as their own learning work.

\subsubsection{RQ2: Do learning supports affect learners' understanding of the CS2 concepts differently?}
Our within-subject comparison of scores in the quizzes provides strong evidence that the GenAI support was less useful for learning the algorithms. This might be partly attributable to the differences that we observed in type of interaction between GenAI and LT (GenAI had few \emph{algorithm understanding} interactions, and no \emph{verifications of understanding}). The pattern of queries with the GenAI might have resulted in more superficial learning and lack of verification of the learners' mental models. This is consistent with metacognitive difficulty results shown by Prather et al. \cite{wideninggap}. 

Participants fared relatively well with AV in terms of scores. This could be because the visual representation was helpful for learning, or because it provided less scaffolding and support, opening more time for actual learning or for the right kind of productive pedagogical friction. Although we do not have definitive evidence, our observations and patterns seem to indicate that the latter is more likely (participants did not perceive the AV as generally useful and the interactions that we observed appear to be mostly ineffective). A future experiment that tests a baseline without any support might clarify this finding.

\subsubsection{RQ3: What do students perceive as the most effective learning support?} 
On aggregate, the LT was perceived as most useful for all aspects except for tracing. In general, AV was rated as least useful (except for tracing), with GenAI generally in between and not preferred on average for any of the aspects. This is despite the larger measured increases in self-efficacy observed with GenAI.

\subsection{Self-efficacy and Metacognition with GenAI}
When looking at the results about RQ1 (self-efficacy) and RQ2 (learning) in conjunction, we find the evidence alarming: GenAI increases student's confidence about their learning more than LT, yet it produces the worst learning results out of the three learning supports, as measured by our quiz. 

This could be interpreted as a failure of metacognition, which has been identified as one of the biggest challenges introduced by GenAI in general~\cite{wideninggap}, and for learners in particular~\cite{metacognitivedifficulties1}. Our analysis indicates that this challenge extends to more advanced CS classes, despite our suspicion that more advanced learners of algorithms would be more mindful about their conceptual learning, as opposed to in introductory programming where it might be easier to forgive students for thinking that the more low-level and mechanical aspects of the programming tasks are less demanding or less worthy of effort. Nevertheless, not everything is a failure of metacognition since, on average, participants seemed to prefer the LT condition for most aspects of the task when they had experienced them all.

\subsection{GenAI vs.\ Live Tutors}
\label{subsec:genaivsLT}
The use of GenAI in education is often motivated as a way to scale personalized assistance and instruction (e.g.,~\cite{CS50_1, AITEachingAssitant, CodeAid}), partly because live tutors (often teaching assistants) are now significantly more expensive than AI. However, the literature is often doubtful that the former can simply replace the latter without sacrificing learning quality, a sentiment which is echoed by students as shown by Bassner et al.~\cite{IrisAITutor}. Our results contribute to the body of explanations as to why. 

We observed different patterns with the two supports: with LT, queries can be quite vague and still successful, since the LT has good context and experience, compared to the proclivity of GenAI to answer a question directly with the potential of misleading the student. This is consistent with Prather et al.'s previous findings~\cite{wideninggap}. Perhaps more importantly, an LT has more latitude to redirect the learner to the right path or to try to build the right model in the mind of the learner, rather than having to strictly answer a question even if that is what the learner wants. 
We partially attribute the behavior pattern differences between GenAI and LT to this.  
Additionally, there are obvious differences in how students interact with other people or systems that surfaced in the study. For example, it would be difficult for a learner to do with an LT the equivalent of just jumping to the code and completely ignoring the explanation like they can with GenAI responses. 
It is not clear whether more advanced LLMs or GenAI-based systems with more context and specifically designed to teach (e.g., in the line of~\cite{IrisAITutor,CodeHelp,kochmar2020automated}) can completely address these issues but, in the meantime we believe that caution is granted.

\subsection{AV and Juggling Visual Representations}
 The AV tool appeared to be difficult for students to use. This was reflected in the the number of interactions which were unsuccessful in progressing the learner in the exercise, in the subjective assessments and rankings, and in our observations of how it was used. Also, the most common interaction with AV tools during the study were primarily passive viewing type interactions, which corresponds to a low level of engagement (\emph{viewing} in Naps et al.'s taxonomy \cite{AVengagement}) which may not be as effective for learning \cite{AVmetastudy}. Participants either did not use or struggled to use the test case creation feature, which would correspond to a higher level of engagement (\textit{changing}). Interestingly, participants often chose to make their own pen and paper visual representations, including in the AV condition. Although this corresponds to a higher level of engagement, participants did not reach this level with the fully-interactive AV tool that we provided.  What obstacles prevented our participants to effectively use the AV tool? Could these obstacles contribute to the lack of general adoption of AV tools?

What we observed leads us to believe that AV might pose a higher barrier than anticipated. In principle, visualization itself is definitely useful. Our experience as instructors has taught us that it is very hard to teach algorithms without graphical representations (even to blind and low vision students \cite{blindstudentscsed, blindtactiledatastructures}). 
Sometimes it is even difficult to communicate one-on-one with students without sketches or a whiteboard. Furthermore, our participants chose to graphically externalize on their own to help with the exercise, unprompted, or went instead searching for the graphical representations in the lecture video. 

Participants' difficulty using the AV tool effectively may have resulted from an insufficient level of instruction on using the tool; learners might not have used the tool effectively if they did not fully understand all the visuals~\cite{rev_understandingvisuals}. We also suspect that an external and dynamic visualization designed by someone else might be too cumbersome to reconcile with the mental representations and the fragile evolving models of the algorithm that learners hold in their minds, especially if they do not have prior experience with the visualization. Additionally, learning how a computer-based visualization works and the subtle but important graphical conventions of the software might just make it appear too cumbersome to use in exchange for what might be perceived as a small missing insight.  

A large corpus of literature in the last forty years has sought to demonstrate the value of interactive visualization in programming instruction and in programming practice (see~\cite{surveyofsuccessfulAV}), and there is a large number of proposed systems (e.g.,~\cite{visualgohalim, pythontutor}). However, some have also wondered why interactive AV is not pervasive and central already for algorithm learning~\cite{AVnotused1,AVnotused2}. We share this surprise and believe that the mental-software representation mismatch hypotheses above could explain our findings and the lack of self adoption of these tools by students. This is also consistent with the findings from Naps et al.~\cite{AVengagement}, which show that visualizations that are built by the student themselves are more likely to be successful. A logical implication is that students need explicit training in visualization to use AV tools effectively, which may perhaps be more broadly applicable to other education aids such as GenAI tutors.

\section{Limitations}
Here we highlight some of the limitations and other caveats of our study, some of which we have mentioned above. First, our sample size is relatively small because our research goals led us to plan a detailed per-participant study that requires several hours of analysis per learner. It is also hard to reach large sample of participants for studies that require this level of involvement (three sessions of more than one hour), even in relatively large cohorts, despite the incentives. Although we were careful to use state-of-the-art statistics to protect us from finding flukes, it will be good to replicate at least some of the specific novel findings with larger cohorts if possible.

Another limitation is that our sessions took place in a separate room, and with a dedicated tutor/experimenter. This deviates from the more common tutorial or lab session with 10-30 students and 1-3 teaching assistants that is common in higher education. Another deviation is that all sessions began with a video to introduce the concept; 
therefore, the results might not be generalizable to classes where students go directly to the lab to learn without having attended or prepared for the material. This can be relatively common if a lab is compulsory but lectures or videos are not, or in flipped-classroom settings that are purposefully designed in this way. As part of the experimental design, we chose to keep the person tutoring the students consistent. We are aware that tutoring quality could vary. We consider the potential consistency of GenAI across students a possible comparative advantage with respect to human tutors.

The study sessions were scheduled to be aligned with the content from the course in which participants were recruited, and as such the topic order of sessions was not counterbalanced across participants. Consequently, our measurements of concept understanding could be influenced by participants developing increased familiarity with CS2 material across later sessions or by fatigue.

We also mentioned that our speculation about the problems with AV and the causes of self-efficacy could be better supported if we had included an additional condition in which participants did not have any support and had to figure out the exercise by themselves. Finally, we acknowledge that the patterns of GenAI use are evolving and more advanced tools might come to widespread use which will require revalidating our results in the future.

\section{Conclusion}
The increasing capability and accessibility of generative AI tools presents opportunities to support learners, but also introduces significant pedagogical challenges for computer science education. Our work generalizes the prior research on GenAI's effects on beginner programmers' self-efficacy and metacognitive awareness by examining how more advanced learners are affected. We show that the previously observed difficulties faced by CS1 students persist among students in CS2, and we extend this prior work by evaluating learning outcomes. We compare GenAI usage with pre-AI supports, namely human tutoring and algorithm visualization. Although self-efficacy is higher when learners use algorithm visualization and GenAI tools compared to when they receive live tutoring, learning outcomes are lower when students use GenAI. Finally, students indicate that they prefer live tutoring over GenAI for learning complex computer science topics. Our findings suggest that GenAI tools are not yet a sufficient direct substitute for traditional learning supports and further highlight the importance of cautious and scaffolded integration of AI in computer science curricula.

\section*{Acknowledgements}
We thank Dr. Rich Little for facilitating this study in parallel with his course offering. We also thank the anonymous reviewers for their detailed and thoughtful feedback, which helped improve the paper. This research was funded by a University of Victoria Fellowship and by NSERC (DG 2020-04401).

\bibliographystyle{ACM-Reference-Format}
\bibliography{sample-base}

@inproceedings{algselfefficacy,
author = {Danielsiek, Holger and Toma, Laura and Vahrenhold, Jan},
title = {An Instrument to Assess Self-Efficacy in Introductory Algorithms Courses},
year = {2017},
isbn = {9781450349680},
publisher = {Association for Computing Machinery},
address = {New York, NY, USA},
url = {https://doi.org/10.1145/3105726.3106171},
doi = {10.1145/3105726.3106171},
booktitle = {Proceedings of the 2017 ACM Conference on International Computing Education Research},
pages = {217–225},
numpages = {9},
location = {Tacoma, Washington, USA},
series = {ICER '17}
}

@inproceedings{robotsarehere,
author = {Prather, James and Denny, Paul and Leinonen, Juho and Becker, Brett A. and Albluwi, Ibrahim and Craig, Michelle and Keuning, Hieke and Kiesler, Natalie and Kohn, Tobias and Luxton-Reilly, Andrew and MacNeil, Stephen and Petersen, Andrew and Pettit, Raymond and Reeves, Brent N. and Savelka, Jaromir},
title = {The Robots Are Here: Navigating the Generative AI Revolution in Computing Education},
year = {2023},
isbn = {9798400704055},
publisher = {Association for Computing Machinery},
address = {New York, NY, USA},
url = {https://doi.org/10.1145/3623762.3633499},
doi = {10.1145/3623762.3633499},
booktitle = {Proceedings of the 2023 Working Group Reports on Innovation and Technology in Computer Science Education},
pages = {108–159},
numpages = {52},
location = {Turku, Finland},
series = {ITiCSE-WGR '23}
}

@inproceedings{robotsarecomingintroprogramming,
author = {Finnie-Ansley, James and Denny, Paul and Becker, Brett A. and Luxton-Reilly, Andrew and Prather, James},
title = {The Robots Are Coming: Exploring the Implications of OpenAI Codex on Introductory Programming},
year = {2022},
isbn = {9781450396431},
publisher = {Association for Computing Machinery},
address = {New York, NY, USA},
url = {https://doi.org/10.1145/3511861.3511863},
doi = {10.1145/3511861.3511863},
booktitle = {Proceedings of the 24th Australasian Computing Education Conference},
pages = {10–19},
numpages = {10},
location = {Virtual Event, Australia},
series = {ACE '22}
}

@inproceedings{GPT4IntroCS,
author = {Savelka, Jaromir and Agarwal, Arav and An, Marshall and Bogart, Chris and Sakr, Majd},
title = {Thrilled by Your Progress! Large Language Models (GPT-4) No Longer Struggle to Pass Assessments in Higher Education Programming Courses},
year = {2023},
isbn = {9781450399760},
publisher = {Association for Computing Machinery},
address = {New York, NY, USA},
url = {https://doi.org/10.1145/3568813.3600142},
doi = {10.1145/3568813.3600142},

booktitle = {Proceedings of the 2023 ACM Conference on International Computing Education Research - Volume 1},
pages = {78–92},
numpages = {15},
location = {Chicago, IL, USA},
series = {ICER '23}
}

@article{llmcriticalthinking,
author = {Jošt, Gregor and Taneski, Viktor and Karakatič, Sašo},
year = {2024},
month = {05},
pages = {4115},
title = {The Impact of Large Language Models on Programming Education and Student Learning Outcomes},
volume = {14},
journal = {Applied Sciences},
doi = {10.3390/app14104115}
}

@inproceedings{CodexCS2,
author = {Finnie-Ansley, James and Denny, Paul and Luxton-Reilly, Andrew and Santos, Eddie Antonio and Prather, James and Becker, Brett A.},
title = {My AI Wants to Know if This Will Be on the Exam: Testing OpenAI’s Codex on CS2 Programming Exercises},
year = {2023},
isbn = {9781450399418},
publisher = {Association for Computing Machinery},
address = {New York, NY, USA},
url = {https://doi.org/10.1145/3576123.3576134},
doi = {10.1145/3576123.3576134},
booktitle = {Proceedings of the 25th Australasian Computing Education Conference},
pages = {97–104},
numpages = {8},
location = {Melbourne, VIC, Australia},
series = {ACE '23}
}

@inproceedings{ComparingCodeExplanations,
author = {Leinonen, Juho and Denny, Paul and MacNeil, Stephen and Sarsa, Sami and Bernstein, Seth and Kim, Joanne and Tran, Andrew and Hellas, Arto},
title = {Comparing Code Explanations Created by Students and Large Language Models},
year = {2023},
isbn = {9798400701382},
publisher = {Association for Computing Machinery},
address = {New York, NY, USA},
url = {https://doi.org/10.1145/3587102.3588785},
doi = {10.1145/3587102.3588785},
booktitle = {Proceedings of the 2023 Conference on Innovation and Technology in Computer Science Education V. 1},
pages = {124–130},
numpages = {7},
location = {Turku, Finland},
series = {ITiCSE 2023}
}

@inproceedings{ProgrammingExercises,
author = {Sarsa, Sami and Denny, Paul and Hellas, Arto and Leinonen, Juho},
title = {Automatic Generation of Programming Exercises and Code Explanations Using Large Language Models},
year = {2022},
isbn = {9781450391948},
publisher = {Association for Computing Machinery},
address = {New York, NY, USA},
url = {https://doi.org/10.1145/3501385.3543957},
doi = {10.1145/3501385.3543957},
booktitle = {Proceedings of the 2022 ACM Conference on International Computing Education Research - Volume 1},
pages = {27–43},
numpages = {17},
location = {Lugano and Virtual Event, Switzerland},
series = {ICER '22}
}

@inproceedings{AITeachingSystematicReview,
author = {Cambaz, Doga and Zhang, Xiaoling},
title = {Use of AI-driven Code Generation Models in Teaching and Learning Programming: a Systematic Literature Review},
year = {2024},
isbn = {9798400704239},
publisher = {Association for Computing Machinery},
address = {New York, NY, USA},
url = {https://doi.org/10.1145/3626252.3630958},
doi = {10.1145/3626252.3630958},
booktitle = {Proceedings of the 55th ACM Technical Symposium on Computer Science Education V. 1},
pages = {172–178},
numpages = {7},
location = {Portland, OR, USA},
series = {SIGCSE 2024}
}

@inproceedings{LLMHelpRequest,
author = {Hellas, Arto and Leinonen, Juho and Sarsa, Sami and Koutcheme, Charles and Kujanp\"{a}\"{a}, Lilja and Sorva, Juha},
title = {Exploring the Responses of Large Language Models to Beginner Programmers’ Help Requests},
year = {2023},
isbn = {9781450399760},
publisher = {Association for Computing Machinery},
address = {New York, NY, USA},
url = {https://doi.org/10.1145/3568813.3600139},
doi = {10.1145/3568813.3600139},
booktitle = {Proceedings of the 2023 ACM Conference on International Computing Education Research - Volume 1},
pages = {93–105},
numpages = {13},
location = {Chicago, IL, USA},
series = {ICER '23}
}

@inproceedings{CS50_2,
author = {Liu, Rongxin and Zhao, Julianna and Xu, Benjamin and Perez, Christopher and Zhukovets, Yuliia and Malan, David J.},
title = {Improving AI in CS50: Leveraging Human Feedback for Better Learning},
year = {2025},
isbn = {9798400705311},
publisher = {Association for Computing Machinery},
address = {New York, NY, USA},
url = {https://doi.org/10.1145/3641554.3701945},
doi = {10.1145/3641554.3701945},
booktitle = {Proceedings of the 56th ACM Technical Symposium on Computer Science Education V. 1},
pages = {715–721},
numpages = {7},
location = {Pittsburgh, PA, USA},
series = {SIGCSETS 2025}
}

@inproceedings{CS50_1,
author = {Liu, Rongxin and Zenke, Carter and Liu, Charlie and Holmes, Andrew and Thornton, Patrick and Malan, David J.},
title = {Teaching CS50 with AI: Leveraging Generative Artificial Intelligence in Computer Science Education},
year = {2024},
isbn = {9798400704239},
publisher = {Association for Computing Machinery},
address = {New York, NY, USA},
url = {https://doi.org/10.1145/3626252.3630938},
doi = {10.1145/3626252.3630938},
booktitle = {Proceedings of the 55th ACM Technical Symposium on Computer Science Education V. 1},
pages = {750–756},
numpages = {7},
location = {Portland, OR, USA},
series = {SIGCSE 2024}
}

@article{CSedinEraofGenAI,
author = {Denny, Paul and Prather, James and Becker, Brett A. and Finnie-Ansley, James and Hellas, Arto and Leinonen, Juho and Luxton-Reilly, Andrew and Reeves, Brent N. and Santos, Eddie Antonio and Sarsa, Sami},
title = {Computing Education in the Era of Generative AI},
year = {2024},
issue_date = {February 2024},
publisher = {Association for Computing Machinery},
address = {New York, NY, USA},
volume = {67},
number = {2},
issn = {0001-0782},
url = {https://doi.org/10.1145/3624720},
doi = {10.1145/3624720},
journal = {Commun. ACM},
month = jan,
pages = {56–67},
numpages = {12}
}

@inproceedings{IrisAITutor,
author = {Bassner, Patrick and Frankford, Eduard and Krusche, Stephan},
title = {Iris: An AI-Driven Virtual Tutor for Computer Science Education},
year = {2024},
isbn = {9798400706004},
publisher = {Association for Computing Machinery},
address = {New York, NY, USA},
url = {https://doi.org/10.1145/3649217.3653543},
doi = {10.1145/3649217.3653543},
booktitle = {Proceedings of the 2024 on Innovation and Technology in Computer Science Education V. 1},
pages = {394–400},
numpages = {7},
location = {Milan, Italy},
series = {ITiCSE 2024}
}

@inproceedings{AITEachingAssitant,
author = {Denny, Paul and MacNeil, Stephen and Savelka, Jaromir and Porter, Leo and Luxton-Reilly, Andrew},
title = {Desirable Characteristics for AI Teaching Assistants in Programming Education},
year = {2024},
isbn = {9798400706004},
publisher = {Association for Computing Machinery},
address = {New York, NY, USA},
url = {https://doi.org/10.1145/3649217.3653574},
doi = {10.1145/3649217.3653574},
booktitle = {Proceedings of the 2024 on Innovation and Technology in Computer Science Education V. 1},
pages = {408–414},
numpages = {7},
location = {Milan, Italy},
series = {ITiCSE 2024}
}

@inproceedings{CodeHelp,
author = {Liffiton, Mark and Sheese, Brad E and Savelka, Jaromir and Denny, Paul},
title = {CodeHelp: Using Large Language Models with Guardrails for Scalable Support in Programming Classes},
year = {2024},
isbn = {9798400716539},
publisher = {Association for Computing Machinery},
address = {New York, NY, USA},
url = {https://doi.org/10.1145/3631802.3631830},
doi = {10.1145/3631802.3631830},
booktitle = {Proceedings of the 23rd Koli Calling International Conference on Computing Education Research},
articleno = {8},
numpages = {11},
location = {Koli, Finland},
series = {Koli Calling '23}
}

@inproceedings{CodeAid,
author = {Kazemitabaar, Majeed and Ye, Runlong and Wang, Xiaoning and Henley, Austin Zachary and Denny, Paul and Craig, Michelle and Grossman, Tovi},
title = {CodeAid: Evaluating a Classroom Deployment of an LLM-based Programming Assistant that Balances Student and Educator Needs},
year = {2024},
isbn = {9798400703300},
publisher = {Association for Computing Machinery},
address = {New York, NY, USA},
url = {https://doi.org/10.1145/3613904.3642773},
doi = {10.1145/3613904.3642773},
booktitle = {Proceedings of the 2024 CHI Conference on Human Factors in Computing Systems},
articleno = {650},
numpages = {20},
location = {Honolulu, HI, USA},
series = {CHI '24}
}

@inproceedings{ProgrammingUsedToBeHard,
author = {Becker, Brett A. and Denny, Paul and Finnie-Ansley, James and Luxton-Reilly, Andrew and Prather, James and Santos, Eddie Antonio},
title = {Programming Is Hard - Or at Least It Used to Be: Educational Opportunities and Challenges of AI Code Generation},
year = {2023},
isbn = {9781450394314},
publisher = {Association for Computing Machinery},
address = {New York, NY, USA},
url = {https://doi.org/10.1145/3545945.3569759},
doi = {10.1145/3545945.3569759},
booktitle = {Proceedings of the 54th ACM Technical Symposium on Computer Science Education V. 1},
pages = {500–506},
numpages = {7},
location = {Toronto ON, Canada},
series = {SIGCSE 2023}
}

@inproceedings{academicintegrity,
author = {Simon and Sheard, Judy and Morgan, Michael and Petersen, Andrew and Settle, Amber and Sinclair, Jane and Cross, Gerry and Riedesel, Charles},
title = {Negotiating the Maze of Academic Integrity in Computing Education},
year = {2016},
isbn = {9781450348829},
publisher = {Association for Computing Machinery},
address = {New York, NY, USA},
url = {https://doi.org/10.1145/3024906.3024910},
doi = {10.1145/3024906.3024910},
booktitle = {Proceedings of the 2016 ITiCSE Working Group Reports},
pages = {57–80},
numpages = {24},
location = {Arequipa, Peru},
series = {ITiCSE '16}
}

@inproceedings{academicintegritycs50,
author = {Malan, David J. and Yu, Brian and Lloyd, Doug},
title = {Teaching Academic Honesty in CS50},
year = {2020},
isbn = {9781450367936},
publisher = {Association for Computing Machinery},
address = {New York, NY, USA},
url = {https://doi.org/10.1145/3328778.3366940},
doi = {10.1145/3328778.3366940},
booktitle = {Proceedings of the 51st ACM Technical Symposium on Computer Science Education},
pages = {282–288},
numpages = {7},
location = {Portland, OR, USA},
series = {SIGCSE '20}
}

@article{academicintegritygenai,
author = {Alier, Marc and García-Peñalvo, Francisco and D. Camba, Jorge},
year = {2024},
month = {03},
pages = {5-14},
title = {Generative Artificial Intelligence in Education: From Deceptive to Disruptive.},
volume = {8},
journal = {International Journal of Interactive Multimedia and Artificial Intelligence},
doi = {10.9781/ijimai.2024.02.011}
}

@inproceedings{overreliance1,
author = {Sheard, Judy and Denny, Paul and Hellas, Arto and Leinonen, Juho and Malmi, Lauri and Simon},
title = {Instructor Perceptions of AI Code Generation Tools - A Multi-Institutional Interview Study},
year = {2024},
isbn = {9798400704239},
publisher = {Association for Computing Machinery},
address = {New York, NY, USA},
url = {https://doi.org/10.1145/3626252.3630880},
doi = {10.1145/3626252.3630880},
pages = {1223–1229},
numpages = {7},
location = {Portland, OR, USA},
series = {SIGCSE 2024}
}

@inproceedings{overreliance2,
author = {Lau, Sam and Guo, Philip},
title = {From "Ban It Till We Understand It" to "Resistance is Futile": How University Programming Instructors Plan to Adapt as More Students Use AI Code Generation and Explanation Tools such as ChatGPT and GitHub Copilot},
year = {2023},
isbn = {9781450399760},
publisher = {Association for Computing Machinery},
address = {New York, NY, USA},
url = {https://doi.org/10.1145/3568813.3600138},
doi = {10.1145/3568813.3600138},
booktitle = {Proceedings of the 2023 ACM Conference on International Computing Education Research - Volume 1},
pages = {106–121},
numpages = {16},
location = {Chicago, IL, USA},
series = {ICER '23}
}

@inproceedings{fewercs2,
author = {Silva, Davi Bernardo and Aguiar, Rafael de Lima and Dvconlo, Diogo Steinke and Silla, Carlos N.},
title = {Recent Studies About Teaching Algorithms (CS1) and Data Structures (CS2) for Computer Science Students},
year = {2019},
publisher = {IEEE Press},
url = {https://doi.org/10.1109/FIE43999.2019.9028702},
doi = {10.1109/FIE43999.2019.9028702},
booktitle = {2019 IEEE Frontiers in Education Conference (FIE)},
pages = {1–8},
numpages = {8},
location = {Covington, KY, USA}
}

@article{focusonintroductoryprogramming,
author = {Deriba, Fitsum and Sanusi, Ismaila and Campbell, Oladele and Oyelere, Solomon},
year = {2024},
month = {01},
pages = {},
title = {Computer Programming Education in the Age of Generative AI: Insights from Empirical Research},
journal = {SSRN Electronic Journal},
doi = {10.2139/ssrn.4891302}
}

@inproceedings{GenAIinupperyear,
author = {Bouvier, Dennis J. and Pereira Cipriano, Bruno and Glassey, Richard and Petrovska, Olga and Anderson, Emma and Birillo, Anastasiia and Dougherty, Ryan and Pettit, Raymond and Pombo, Nuno and Rahimi, Ebrahim and Ramakrishnan, Charanya and Steinmaurer, Alexander and Taneja, Shubbhi and Usman, Muhammad and Vadaparty, Annapurna},
title = {The Rest of the Robots: Generative AI in Post-introductory Computing Education},
year = {2026},
isbn = {9798400721670},
publisher = {Association for Computing Machinery},
address = {New York, NY, USA},
url = {https://doi.org/10.1145/3760545.3783970},
doi = {10.1145/3760545.3783970},
booktitle = {Proceedings of the 2025 Working Group Reports on Innovation and Technology in Computer Science Education},
pages = {61–107},
numpages = {47},
location = {Netherlands},
series = {ITiCSE-WGR 2025}
}

@inproceedings{interactionchallengesnguyen,
author = {Nguyen, Sydney and Babe, Hannah McLean and Zi, Yangtian and Guha, Arjun and Anderson, Carolyn Jane and Feldman, Molly Q},
title = {How Beginning Programmers and Code LLMs (Mis)read Each Other},
year = {2024},
isbn = {9798400703300},
publisher = {Association for Computing Machinery},
address = {New York, NY, USA},
url = {https://doi.org/10.1145/3613904.3642706},
doi = {10.1145/3613904.3642706},
booktitle = {Proceedings of the 2024 CHI Conference on Human Factors in Computing Systems},
articleno = {651},
numpages = {26},
location = {Honolulu, HI, USA},
series = {CHI '24}
}

@article{interactionchallengesprather,
author = {Prather, James and Reeves, Brent N. and Denny, Paul and Becker, Brett A. and Leinonen, Juho and Luxton-Reilly, Andrew and Powell, Garrett and Finnie-Ansley, James and Santos, Eddie Antonio},
title = {“It’s Weird That it Knows What I Want”: Usability and Interactions with Copilot for Novice Programmers},
year = {2023},
issue_date = {February 2024},
publisher = {Association for Computing Machinery},
address = {New York, NY, USA},
volume = {31},
number = {1},
issn = {1073-0516},
url = {https://doi.org/10.1145/3617367},
doi = {10.1145/3617367},
journal = {ACM Trans. Comput.-Hum. Interact.},
month = nov,
articleno = {4},
numpages = {31}
}

@inproceedings{interactionscodetracing,
author = {Adeeb, Elmira and Muldner, Kasia},
title = {How Do Novice Programmers Solve Code-Tracing Problems When ChatGPT Is Available? A Qualitative Analysis.},
year = {2025},
isbn = {9798400713408},
publisher = {Association for Computing Machinery},
address = {New York, NY, USA},
url = {https://doi.org/10.1145/3702652.3744207},
doi = {10.1145/3702652.3744207},
booktitle = {Proceedings of the 2025 ACM Conference on International Computing Education Research V.1},
pages = {421–434},
numpages = {14},
location = {
},
series = {ICER '25}
}

@inproceedings{interactionspythoncourse,
author = {Kazemitabaar, Majeed and Hou, Xinying and Henley, Austin and Ericson, Barbara Jane and Weintrop, David and Grossman, Tovi},
title = {How Novices Use LLM-based Code Generators to Solve CS1 Coding Tasks in a Self-Paced Learning Environment},
year = {2024},
isbn = {9798400716539},
publisher = {Association for Computing Machinery},
address = {New York, NY, USA},
url = {https://doi.org/10.1145/3631802.3631806},
doi = {10.1145/3631802.3631806},
booktitle = {Proceedings of the 23rd Koli Calling International Conference on Computing Education Research},
articleno = {3},
numpages = {12},
location = {Koli, Finland},
series = {Koli Calling '23}
}

@article{selfefficacydefinition,
title = {Self-efficacy: Toward a unifying theory of behavioral change},
journal = {Advances in Behaviour Research and Therapy},
volume = {1},
number = {4},
pages = {139-161},
year = {1978},
note = {Perceived Self-Efficacy: Analyses of Bandura's Theory of Behavioural Change},
issn = {0146-6402},
doi = {https://doi.org/10.1016/0146-6402(78)90002-4},
url = {https://www.sciencedirect.com/science/article/pii/0146640278900024},
author = {Albert Bandura},
}

@inproceedings{selfefficacypredictor1,
author = {Kinnunen, P\"{a}ivi and Simon, Beth},
title = {CS majors' self-efficacy perceptions in CS1: results in light of social cognitive theory},
year = {2011},
isbn = {9781450308298},
publisher = {Association for Computing Machinery},
address = {New York, NY, USA},
url = {https://doi.org/10.1145/2016911.2016917},
doi = {10.1145/2016911.2016917},
booktitle = {Proceedings of the Seventh International Workshop on Computing Education Research},
pages = {19–26},
numpages = {8},
location = {Providence, Rhode Island, USA},
series = {ICER '11}
}

@article{selfefficacypredictor2,
author = {Lishinski, Alex and Yadav, Aman},
title = {Self-evaluation Interventions: Impact on Self-efficacy and Performance in Introductory Programming},
year = {2021},
issue_date = {September 2021},
publisher = {Association for Computing Machinery},
address = {New York, NY, USA},
volume = {21},
number = {3},
url = {https://doi.org/10.1145/3447378},
doi = {10.1145/3447378},
journal = {ACM Trans. Comput. Educ.},
month = jun,
articleno = {23},
numpages = {28}
}

@inproceedings{studentAIinteraction,
author = {Amoozadeh, Matin and Nam, Daye and Prol, Daniel and Alfageeh, Ali and Prather, James and Hilton, Michael and Srinivasa Ragavan, Sruti and Alipour, Amin},
title = {Student-AI Interaction: A Case Study of CS1 students},
year = {2024},
isbn = {9798400710384},
publisher = {Association for Computing Machinery},
address = {New York, NY, USA},
url = {https://doi.org/10.1145/3699538.3699567},
doi = {10.1145/3699538.3699567},
booktitle = {Proceedings of the 24th Koli Calling International Conference on Computing Education Research},
articleno = {13},
numpages = {13},
location = {
},
series = {Koli Calling '24}
}

@inproceedings{selfregulationselfefficacy,
author = {Margulieux, Lauren E. and Prather, James and Reeves, Brent N. and Becker, Brett A. and Cetin Uzun, Gozde and Loksa, Dastyni and Leinonen, Juho and Denny, Paul},
title = {Self-Regulation, Self-Efficacy, and Fear of Failure Interactions with How Novices Use LLMs to Solve Programming Problems},
year = {2024},
isbn = {9798400706004},
publisher = {Association for Computing Machinery},
address = {New York, NY, USA},
url = {https://doi.org/10.1145/3649217.3653621},
doi = {10.1145/3649217.3653621},
booktitle = {Proceedings of the 2024 on Innovation and Technology in Computer Science Education V. 1},
pages = {276–282},
numpages = {7},
location = {Milan, Italy},
series = {ITiCSE 2024}
}

@incollection{metacognitiondefinition,
  title={Metacognitive aspects of problem solving},
  author={Flavell, John H},
  booktitle={The nature of intelligence},
  pages={231--236},
  year={2024},
  publisher={Routledge}
}

@inproceedings{metacognitioninprogramming,
author = {Prather, James and Becker, Brett A. and Craig, Michelle and Denny, Paul and Loksa, Dastyni and Margulieux, Lauren},
title = {What Do We Think We Think We Are Doing? Metacognition and Self-Regulation in Programming},
year = {2020},
isbn = {9781450370929},
publisher = {Association for Computing Machinery},
address = {New York, NY, USA},
url = {https://doi.org/10.1145/3372782.3406263},
doi = {10.1145/3372782.3406263},
booktitle = {Proceedings of the 2020 ACM Conference on International Computing Education Research},
pages = {2–13},
numpages = {12},
location = {Virtual Event, New Zealand},
series = {ICER '20}
}

@inproceedings{metacognitivedifficulties1,
author = {Prather, James and Pettit, Raymond and McMurry, Kayla and Peters, Alani and Homer, John and Cohen, Maxine},
title = {Metacognitive Difficulties Faced by Novice Programmers in Automated Assessment Tools},
year = {2018},
isbn = {9781450356282},
publisher = {Association for Computing Machinery},
address = {New York, NY, USA},
url = {https://doi.org/10.1145/3230977.3230981},
doi = {10.1145/3230977.3230981},
booktitle = {Proceedings of the 2018 ACM Conference on International Computing Education Research},
pages = {41–50},
numpages = {10},
location = {Espoo, Finland},
series = {ICER '18}
}

@inproceedings{wideninggap,
author = {Prather, James and Reeves, Brent N and Leinonen, Juho and MacNeil, Stephen and Randrianasolo, Arisoa S and Becker, Brett A. and Kimmel, Bailey and Wright, Jared and Briggs, Ben},
title = {The Widening Gap: The Benefits and Harms of Generative AI for Novice Programmers},
year = {2024},
isbn = {9798400704758},
publisher = {Association for Computing Machinery},
address = {New York, NY, USA},
url = {https://doi.org/10.1145/3632620.3671116},
doi = {10.1145/3632620.3671116},
booktitle = {Proceedings of the 2024 ACM Conference on International Computing Education Research - Volume 1},
pages = {469–486},
numpages = {18},
location = {Melbourne, VIC, Australia},
series = {ICER '24}
}

@article{MSQL,
  title={A manual for the use of the Motivated Strategies for Learning Questionnaire (MSLQ).},
  author={Pintrich, Paul R and others},
  year={1991},
  publisher={ERIC}
}

@article{CS1selfefficacysurvey,
  title={Development and validation of scores on a computer programming self-efficacy scale and group analyses of novice programmer self-efficacy},
  author={Ramalingam, Vennila and Wiedenbeck, Susan},
  journal={Journal of Educational Computing Research},
  volume={19},
  number={4},
  pages={367--381},
  year={1998},
  publisher={SAGE Publications Sage CA: Los Angeles, CA}
}

@article{oldAV,
author = {Brown, Marc H. and Sedgewick, Robert},
title = {A system for algorithm animation},
year = {1984},
issue_date = {July 1984},
publisher = {Association for Computing Machinery},
address = {New York, NY, USA},
volume = {18},
number = {3},
issn = {0097-8930},
url = {https://doi.org/10.1145/964965.808596},
doi = {10.1145/964965.808596},
journal = {SIGGRAPH Comput. Graph.},
month = jan,
pages = {177–186},
numpages = {10}
}

@article{AVstateofthefield,
author = {Shaffer, Clifford A. and Cooper, Matthew L. and Alon, Alexander Joel D. and Akbar, Monika and Stewart, Michael and Ponce, Sean and Edwards, Stephen H.},
title = {Algorithm Visualization: The State of the Field},
year = {2010},
issue_date = {August 2010},
publisher = {Association for Computing Machinery},
address = {New York, NY, USA},
volume = {10},
number = {3},
url = {https://doi.org/10.1145/1821996.1821997},
doi = {10.1145/1821996.1821997},
journal = {ACM Trans. Comput. Educ.},
month = aug,
articleno = {9},
numpages = {22}
}

@article{surveyofsuccessfulAV,
author = {Urquiza-Fuentes, Jaime and Vel\'{a}zquez-Iturbide, J. \'{A}ngel},
title = {A Survey of Successful Evaluations of Program Visualization and Algorithm Animation Systems},
year = {2009},
issue_date = {June 2009},
publisher = {Association for Computing Machinery},
address = {New York, NY, USA},
volume = {9},
number = {2},
url = {https://doi.org/10.1145/1538234.1538236},
doi = {10.1145/1538234.1538236},
journal = {ACM Trans. Comput. Educ.},
month = jun,
articleno = {9},
numpages = {21}
}

@article{visualgohalim,
  title={Visualgo--visualising data structures and algorithms through animation},
  author={Halim, Steven},
  journal={Olympiads in informatics},
  volume={9},
  pages={243--245},
  year={2015}
}

@article{AP-ASD1,
  title={AP-ASD1: An Indonesian Desktop-based Educational Tool for Basic Data Structure Course},
  author={Christiawan, Lucky and Karnalim, Oscar},
  journal={Jurnal Teknik Informatika dan Sistem Informasi},
  volume={2},
  number={1},
  year={2016}
}

@article{jhave,
  title={Jhav{\'e}: Supporting algorithm visualization},
  author={Naps, Thomas L},
  journal={IEEE Computer Graphics and Applications},
  volume={25},
  number={5},
  pages={49--55},
  year={2005},
  publisher={IEEE}
}

@article{vizalgo,
  title={Algorithm visualization using the vizalgo platform},
  author={Simon{\'a}k, Slavom{\'\i}r},
  journal={Acta Electrotechnica et Informatica},
  volume={13},
  number={2},
  pages={54},
  year={2013},
  publisher={De Gruyter Poland}
}

@inproceedings{opendsa,
  title={OpenDSA: beginning a community active-ebook project},
  author={Shaffer, Clifford A and Karavirta, Ville and Korhonen, Ari and Naps, Thomas L},
  booktitle={Proceedings of the 11th Koli Calling International Conference on computing education research},
  pages={112--117},
  year={2011}
}

@inproceedings{pythontutor,
  title={Online python tutor: embeddable web-based program visualization for cs education},
  author={Guo, Philip J},
  booktitle={Proceeding of the 44th ACM technical symposium on Computer science education},
  pages={579--584},
  year={2013}
}

@inproceedings{pinti,
  title={P-Inti: Interactive visual representation of programming concepts for learning and instruction},
  author={Halaharvi, Shishir and M{\'e}ndez, Gonzalo Gabriel and Mansoor, Hamid and Yong, Quinton and Milani, Alessandra Maciel Paz and Storey, Margaret-Anne and Nacenta, Miguel A},
  booktitle={2024 IEEE Symposium on Visual Languages and Human-Centric Computing (VL/HCC)},
  pages={199--210},
  year={2024},
  organization={IEEE}
}

@article{programvisualization,
  title={Visual programming and program visualization towards an ideal visual software engineering system},
  author={Bentrad, Sassi and Meslati, Djamel and others},
  journal={ACEEE International Journal on Information Technology},
  volume={1},
  number={3},
  pages={43--49},
  year={2011},
  publisher={Citeseer}
}

@inproceedings{jeliot,
  title={Visualizing programs with Jeliot 3},
  author={Moreno, Andr{\'e}s and Myller, Niko and Sutinen, Erkki and Ben-Ari, Mordechai},
  booktitle={Proceedings of the working conference on Advanced visual interfaces},
  pages={373--376},
  year={2004}
}

@inproceedings{ville,
  title={VILLE: a language-independent program visualization tool},
  author={Rajala, Teemu and Laakso, Mikko-Jussi and Kaila, Erkki and Salakoski, Tapio},
  booktitle={Proceedings of the Seventh Baltic Sea Conference on Computing Education Research-Volume 88},
  pages={151--159},
  year={2007}
}

@article{AVmetastudy,
  title={A meta-study of algorithm visualization effectiveness},
  author={Hundhausen, Christopher D and Douglas, Sarah A and Stasko, John T},
  journal={Journal of Visual Languages \& Computing},
  volume={13},
  number={3},
  pages={259--290},
  year={2002},
  publisher={Elsevier}
}

@inproceedings{AVengagement,
author = {Naps, Thomas L. and R\"{o}\ss{}ling, Guido and Almstrum, Vicki and Dann, Wanda and Fleischer, Rudolf and Hundhausen, Chris and Korhonen, Ari and Malmi, Lauri and McNally, Myles and Rodger, Susan and Vel\'{a}zquez-Iturbide, J. \'{A}ngel},
title = {Exploring the role of visualization and engagement in computer science education},
year = {2002},
isbn = {9781450374491},
publisher = {Association for Computing Machinery},
address = {New York, NY, USA},
url = {https://doi.org/10.1145/960568.782998},
doi = {10.1145/960568.782998},
booktitle = {Working Group Reports from ITiCSE on Innovation and Technology in Computer Science Education},
pages = {131–152},
numpages = {22},
location = {Aarhus, Denmark},
series = {ITiCSE-WGR '02}
}

@article{avcognitiveload1,
  title={Measuring learners’ cognitive load when engaged with an algorithm visualization tool},
  author={Fathi, Razieh and Teresco, James D and Regan, Kenneth},
  journal={Journal of Information Systems Applied Research},
  volume={16},
  number={3},
  year={2023}
}

@inproceedings{avcognitiveload2,
  title={Using visualization to reduce the cognitive load of threshold concepts in computer programming},
  author={Winter, Victor and Friend, Michelle and Matthews, Michael and Love, Betty and Vasireddy, Sanghamithra},
  booktitle={2019 IEEE Frontiers in Education Conference (FIE)},
  pages={1--9},
  year={2019},
  organization={IEEE}
}

@article{AVselfefficacy,
  title={Measuring effectiveness of graph visualizations: A cognitive load perspective},
  author={Huang, Weidong and Eades, Peter and Hong, Seok-Hee},
  journal={Information Visualization},
  volume={8},
  number={3},
  pages={139--152},
  year={2009},
  publisher={SAGE Publications Sage UK: London, England}
}

@inproceedings{AVnotused2,
author = {Isohanni, Essi and J\"{a}rvinen, Hannu-Matti},
title = {Are visualization tools used in programming education? by whom, how, why, and why not?},
year = {2014},
isbn = {9781450330657},
publisher = {Association for Computing Machinery},
address = {New York, NY, USA},
url = {https://doi.org/10.1145/2674683.2674688},
doi = {10.1145/2674683.2674688},
booktitle = {Proceedings of the 14th Koli Calling International Conference on Computing Education Research},
pages = {35–40},
numpages = {6},
location = {Koli, Finland},
series = {Koli Calling '14}
}

@inproceedings{AVnotused1,
author = {Knobelsdorf, Maria and Isohanni, Essi and Tenenberg, Josh},
title = {The reasons might be different: why students and teachers do not use visualization tools},
year = {2012},
isbn = {9781450317955},
publisher = {Association for Computing Machinery},
address = {New York, NY, USA},
url = {https://doi.org/10.1145/2401796.2401797},
doi = {10.1145/2401796.2401797},
booktitle = {Proceedings of the 12th Koli Calling International Conference on Computing Education Research},
pages = {1–10},
numpages = {10},
location = {Koli, Finland},
series = {Koli Calling '12}
}

@article{AVeffectiveness,
  title={An empirical study on factors influencing the effectiveness of algorithm visualization},
  author={Lazaridis, Vassilios and Samaras, Nikolaos and Sifaleras, Angelo},
  journal={Computer Applications in Engineering Education},
  volume={21},
  number={3},
  pages={410--420},
  year={2013},
  publisher={Wiley Online Library}
}

@inproceedings{AAVeffectiveness,
author = {Farghally, Mohammed F. and Koh, Kyu Han and Shahin, Hossameldin and Shaffer, Clifford A.},
title = {Evaluating the Effectiveness of Algorithm Analysis Visualizations},
year = {2017},
isbn = {9781450346986},
publisher = {Association for Computing Machinery},
address = {New York, NY, USA},
url = {https://doi.org/10.1145/3017680.3017698},
doi = {10.1145/3017680.3017698},
booktitle = {Proceedings of the 2017 ACM SIGCSE Technical Symposium on Computer Science Education},
pages = {201–206},
numpages = {6},
location = {Seattle, Washington, USA},
series = {SIGCSE '17}
}

@article{kruschke_bayesian_2021,
    title = {Bayesian {Analysis} {Reporting} {Guidelines}},
    volume = {5},
    copyright = {2021 The Author(s)},
    issn = {2397-3374},
    url = {https://www.nature.com/articles/s41562-021-01177-7},
    doi = {10.1038/s41562-021-01177-7},
    language = {en},
    number = {10},
    urldate = {2022-08-03},
    journal = {Nature Human Behaviour},
    publisher = {Nature Publishing Group},
    author = {Kruschke, John K.},
    month = oct,
    year = {2021},
    note = {Number: 10},
    pages = {1282--1291},
}

@Manual{Rlanguage,
    title = {R: A Language and Environment for Statistical Computing},
    author = {{R Core Team}},
    organization = {R Foundation for Statistical Computing},
    address = {Vienna, Austria},
    year = {2025},
    url = {https://www.R-project.org/},
  }

@book{kline_beyond_2013,
    address = {Washington, DC, US},
    series = {Beyond significance testing: {Statistics} reform in the behavioral sciences, 2nd ed},
    title = {Beyond significance testing: {Statistics} reform in the behavioral sciences, 2nd ed},
    isbn = {978-1-4338-1278-1},
    shorttitle = {Beyond significance testing},
    doi = {10.1037/14136-000},
    publisher = {American Psychological Association},
    author = {Kline, Rex B.},
    year = {2013},
    note = {Pages: xi, 349},
}

@article{cumming_new_2014,
    title = {The {New} {Statistics} {Why} and {How}},
    volume = {25},
    issn = {0956-7976, 1467-9280},
    url = {http://pss.sagepub.com/content/25/1/7},
    doi = {10.1177/0956797613504966},
    language = {en},
    number = {1},
    urldate = {2015-08-04},
    journal = {Psychological Science},
    author = {Cumming, Geoff},
    month = jan,
    year = {2014},
    pages = {7--29},
}

@article{rjags,
  title={Doing Bayesian data analysis: A tutorial with R, JAGS, and Stan},
  author={Kruschke, John},
  year={2014},
  publisher={Academic Press}
}

@inproceedings{plummer2003jags,
  title={JAGS: A program for analysis of Bayesian graphical models using Gibbs sampling},
  author={Plummer, Martyn and others},
  booktitle={Proceedings of the 3rd international workshop on distributed statistical computing},
  volume={124},
  number={125.10},
  pages={1--10},
  year={2003},
  organization={Vienna, Austria}
}

@article{chatgptThreats,
  title={ChatGPT for education and research: Opportunities, threats, and strategies},
  author={Rahman, Md Mostafizer and Watanobe, Yutaka},
  journal={Applied sciences},
  volume={13},
  number={9},
  pages={5783},
  year={2023},
  publisher={MDPI}
}

@article{genAIscaffolding,
  title={Generative AI in introductory programming},
  author={Becker, Brett A and Craig, Michelle and Denny, Paul and Keuning, Hieke and Kiesler, Natalie and Leinonen, Juho and Luxton-Reilly, Andrew and Prather, James and Quille, Keith},
  journal={Computer Science Curricula},
  pages={438--439},
  year={2023}
}

@article{cs2difficulties,
  title={Difficulties in learning the data structures course: Literature review},
  author={Mtaho, Adam Basigie and Mselle, Leonard James},
  journal={The Journal of Informatics},
  volume={4},
  number={1},
  pages={26--55},
  year={2024}
}

@inproceedings{cs2dropoutrate,
  title={A course on algorithms and data structures using on-line judging},
  author={G{\'a}rcia-Mateos, Gin{\'e}s and Fern{\'a}ndez-Alem{\'a}n, Jos{\'e} Luis},
  booktitle={Proceedings of the 14th annual ACM SIGCSE conference on innovation and technology in computer science education},
  pages={45--49},
  year={2009}
}

@inproceedings{barriersinproblemsolving,
author = {Loksa, Dastyni and Ko, Amy J. and Jernigan, Will and Oleson, Alannah and Mendez, Christopher J. and Burnett, Margaret M.},
title = {Programming, Problem Solving, and Self-Awareness: Effects of Explicit Guidance},
year = {2016},
isbn = {9781450333627},
publisher = {Association for Computing Machinery},
address = {New York, NY, USA},
url = {https://doi.org/10.1145/2858036.2858252},
doi = {10.1145/2858036.2858252},
booktitle = {Proceedings of the 2016 CHI Conference on Human Factors in Computing Systems},
pages = {1449–1461},
numpages = {13},
location = {San Jose, California, USA},
series = {CHI '16}
}

@article{thematicanalysis,
  title={Thematic analysis},
  author={Clarke, Victoria and Braun, Virginia},
  journal={The journal of positive psychology},
  volume={12},
  number={3},
  pages={297--298},
  year={2017},
  publisher={Taylor \& Francis}
}

@article{harking,
  title={HARKing: Hypothesizing after the results are known},
  author={Kerr, Norbert L},
  journal={Personality and social psychology review},
  volume={2},
  number={3},
  pages={196--217},
  year={1998},
  publisher={Sage Publications Sage CA: Los Angeles, CA}
}

@article{bloomstaxonomy,
  title={A revision of Bloom's taxonomy: An overview},
  author={Krathwohl, David R},
  journal={Theory into practice},
  volume={41},
  number={4},
  pages={212--218},
  year={2002},
  publisher={Taylor \& Francis}
}

@inproceedings{bloomstaxonomycs,
author = {Thompson, Errol and Luxton-Reilly, Andrew and Whalley, Jacqueline L. and Hu, Minjie and Robbins, Phil},
title = {Bloom's taxonomy for CS assessment},
year = {2008},
isbn = {9781920682590},
publisher = {Australian Computer Society, Inc.},
address = {AUS},
booktitle = {Proceedings of the Tenth Conference on Australasian Computing Education - Volume 78},
pages = {155–161},
numpages = {7},
location = {Wollongong, NSW, Australia},
series = {ACE '08}
}

@phdthesis{blindstudentscsed,
  title={Understanding and Improving Blind Students’ Access to Visual Information in Computer Science Education},
  author={Baker, Catherine Marie},
  year={2017}
}

@inproceedings{blindtactiledatastructures,
  title={Improving understanding of data structures for the blind with tactile media and a user-centered iterative approach},
  author={Crockett, April R and Gannod, Gerald C},
  booktitle={2020 IEEE Frontiers in Education Conference (FIE)},
  pages={1--8},
  year={2020},
  organization={IEEE}
}

@inproceedings{kochmar2020automated,
  title={Automated personalized feedback improves learning gains in an intelligent tutoring system},
  author={Kochmar, Ekaterina and Vu, Dung Do and Belfer, Robert and Gupta, Varun and Serban, Iulian Vlad and Pineau, Joelle},
  booktitle={International conference on artificial intelligence in education},
  pages={140--146},
  year={2020},
  organization={Springer}
}

@article{rev_understandingvisuals,
author = {Petre, Marian},
title = {Why looking isn't always seeing: readership skills and graphical programming},
year = {1995},
issue_date = {June 1995},
publisher = {Association for Computing Machinery},
address = {New York, NY, USA},
volume = {38},
number = {6},
issn = {0001-0782},
url = {https://doi.org/10.1145/203241.203251},
doi = {10.1145/203241.203251},
journal = {Commun. ACM},
month = jun,
pages = {33–44},
numpages = {12}
}

@inproceedings{rev_genAICS1learningoutcomes,
author = {Vadaparty, Annapurna and Smith, David H., IV and Srinath, Samvrit and Padala, Mounika and Alvarado, Christine and Gorson Benario, Jamie and Porter, Leo and Zingaro, Daniel},
title = {Evaluating CS1-LLM: Integrating LLMs and Examining Student Outcomes in an Introductory Computer Science Course},
year = {2026},
isbn = {9798400723520},
publisher = {Association for Computing Machinery},
address = {New York, NY, USA},
url = {https://doi.org/10.1145/3786228.3786234},
doi = {10.1145/3786228.3786234},
booktitle = {Proceedings of the 28th Australasian Computing Education Conference},
pages = {32–41},
numpages = {10},
location = {
},
series = {ACE '26}
}

@inproceedings{rev_CS2inflection1,
author = {Hooshangi, Sara and Ellis, Margaret and Edwards, Stephen H.},
title = {Factors Influencing Student Performance and Persistence in CS2},
year = {2022},
isbn = {9781450390705},
publisher = {Association for Computing Machinery},
address = {New York, NY, USA},
url = {https://doi.org/10.1145/3478431.3499272},
doi = {10.1145/3478431.3499272},
booktitle = {Proceedings of the 53rd ACM Technical Symposium on Computer Science Education - Volume 1},
pages = {286–292},
numpages = {7},
location = {Providence, RI, USA},
series = {SIGCSE 2022}
}

@inproceedings{rev_CS2inflection2,
author = {Horton, Diane and Craig, Michelle},
title = {Drop, Fail, Pass, Continue: Persistence in CS1 and Beyond in Traditional and Inverted Delivery},
year = {2015},
isbn = {9781450329668},
publisher = {Association for Computing Machinery},
address = {New York, NY, USA},
url = {https://doi.org/10.1145/2676723.2677273},
doi = {10.1145/2676723.2677273},
booktitle = {Proceedings of the 46th ACM Technical Symposium on Computer Science Education},
pages = {235–240},
numpages = {6},
location = {Kansas City, Missouri, USA},
series = {SIGCSE '15}
}

@inproceedings{rev_CS2inflection3,
author = {Layman, Lucas and Song, Yang and Guinn, Curry},
title = {Toward Predicting Success and Failure in CS2: A Mixed-Method Analysis},
year = {2020},
isbn = {9781450371056},
publisher = {Association for Computing Machinery},
address = {New York, NY, USA},
url = {https://doi.org/10.1145/3374135.3385277},
doi = {10.1145/3374135.3385277},
booktitle = {Proceedings of the 2020 ACM Southeast Conference},
pages = {218–225},
numpages = {8},
location = {Tampa, FL, USA},
series = {ACMSE '20}
}

@inproceedings{rev_AIgrading,
author = {Zhao, Chenyan and Silva, Mariana and Poulsen, Seth},
title = {Language Models are Few-Shot Graders},
year = {2025},
isbn = {978-3-031-98458-7},
publisher = {Springer-Verlag},
address = {Berlin, Heidelberg},
url = {https://doi.org/10.1007/978-3-031-98459-4_1},
doi = {10.1007/978-3-031-98459-4_1},
booktitle = {Artificial Intelligence in Education: 26th International Conference, AIED 2025, Palermo, Italy, July 22–26, 2025, Proceedings, Part IV},
pages = {3–16},
numpages = {14},
location = {Palermo, Italy}
}

@inproceedings{rev_genAItutordialogue,
author = {Scarlatos, Alexander and Liu, Naiming and Lee, Jaewook and Baraniuk, Richard and Lan, Andrew},
title = {Training LLM-Based Tutors to Improve Student Learning Outcomes in Dialogues},
year = {2025},
isbn = {978-3-031-98413-6},
publisher = {Springer-Verlag},
address = {Berlin, Heidelberg},
url = {https://doi.org/10.1007/978-3-031-98414-3_18},
doi = {10.1007/978-3-031-98414-3_18},
booktitle = {Artificial Intelligence in Education: 26th International Conference, AIED 2025, Palermo, Italy, July 22–26, 2025, Proceedings, Part I},
pages = {251–266},
numpages = {16},
location = {Palermo, Italy}
}

@misc{rev_genAItutoraccessible,
      title={Leveraging LLM Tutoring Systems for Non-Native English Speakers in Introductory CS Courses}, 
      author={Ismael Villegas Molina and Audria Montalvo and Benjamin Ochoa and Paul Denny and Leo Porter},
      year={2025},
      eprint={2411.02725},
      archivePrefix={arXiv},
      primaryClass={cs.HC},
      url={https://arxiv.org/abs/2411.02725}, 
}

@INPROCEEDINGS{rev_genAIalgscourse,
  author={Palacios-Alonso, Daniel and Urquiza-Fuentes, Jaime and Velázquez-Iturbide, J. Ángel and Guillén-García, Julio},
  booktitle={2024 IEEE Global Engineering Education Conference (EDUCON)}, 
  title={Experiences and Proposals of Use of Generative AI in Advanced Software Courses}, 
  year={2024},
  volume={},
  number={},
  pages={1-10},
  doi={10.1109/EDUCON60312.2024.10578869}}

@article{rev_PVsurvey,
author = {Sorva, Juha and Karavirta, Ville and Malmi, Lauri},
title = {A Review of Generic Program Visualization Systems for Introductory Programming Education},
year = {2013},
issue_date = {November 2013},
publisher = {Association for Computing Machinery},
address = {New York, NY, USA},
volume = {13},
number = {4},
url = {https://doi.org/10.1145/2490822},
doi = {10.1145/2490822},
journal = {ACM Trans. Comput. Educ.},
month = nov,
articleno = {15},
numpages = {64}
}

@article{rev_PVinalgscourse,
author = {Vel\'{a}zquez-Iturbide, J. \'{A}ngel and P\'{e}rez-Carrasco, Antonio},
title = {How to use the SRec visualization system in programming and algorithm courses},
year = {2016},
issue_date = {September 2016},
publisher = {Association for Computing Machinery},
address = {New York, NY, USA},
volume = {7},
number = {3},
issn = {2153-2184},
url = {https://doi.org/10.1145/2948070},
doi = {10.1145/2948070},
journal = {ACM Inroads},
month = aug,
pages = {42–49},
numpages = {8}
}

\clearpage
\appendix
\onecolumn
\section{Tables of Quantitative Statistical Tests}

\begin{table*}[!ht]
\centering
\caption{Statistical parameters and tests of change in self-efficacy and quiz scores.}

\renewcommand{\arraystretch}{1.2}

\resizebox{\textwidth}{!}{
\begin{tabular}{|c|c|c|c|c|c|c|c|}
\Xhline{2\arrayrulewidth}

\textbf{Measurement} & \textbf{Test} & \textbf{Parameter} & \textbf{Probability} & \textbf{Mean} & \textbf{Median} & $\boldsymbol{\sigma}$ & \textbf{HDI} \\
\Xhline{2\arrayrulewidth}

\multirow{6}{*}{\centering Change in SE}
& LT & $\beta_0[\mathrm{LT}] + \beta_1[\mathrm{LT}]$ & -- & 6.52 & 6.52 & 1.81 & [3.00, 10.22] \\
\cline{2-8}

& GenAI & $\beta_0[\mathrm{GenAI}] + \beta_1[\mathrm{GenAI}]$ & -- & 8.74 & 8.74 & 1.71 & [5.40, 12.15] \\
\cline{2-8}

& AV & $\beta_0[\mathrm{AV}] + \beta_1[\mathrm{AV}]$ & -- & 10.48 & 10.49 & 1.70 & [7.10, 13.83] \\
\cline{2-8}

& LT $>$ GenAI & $\beta_1[\mathrm{LT}] - \beta_1[\mathrm{GenAI}]$ & 0.18 & -2.22 & -2.21 & 2.50 & [-7.19, 2.72] \\
\cline{2-8}

& LT $>$ AV & $\beta_1[\mathrm{LT}] - \beta_1[\mathrm{AV}]$ & 0.05 & -3.96 & -3.95 & 2.50 & [-8.92, 0.87] \\
\cline{2-8}

& AV $>$ GenAI & $\beta_1[\mathrm{AV}] - \beta_1[\mathrm{GenAI}]$ & 0.78 & 1.74 & 1.75 & 2.41 & [-3.11, 6.45] \\
\Xhline{2\arrayrulewidth}

\multirow{6}{*}{\centering Quiz Score}
& LT & $\beta_0[\mathrm{LT}] + \beta_1[\mathrm{LT}]$ & -- & 5.55 & 5.55 & 0.43 & [4.70, 6.39] \\
\cline{2-8}

& GenAI & $\beta_0[\mathrm{GenAI}] + \beta_1[\mathrm{GenAI}]$ & -- & 4.42 & 4.42 & 0.40 & [3.63, 5.23] \\
\cline{2-8}

& AV & $\beta_0[\mathrm{AV}] + \beta_1[\mathrm{AV}]$ & -- & 5.42 & 5.41 & 0.40 & [4.64, 6.23] \\
\cline{2-8}

& LT $>$ GenAI & $\beta_1[\mathrm{LT}] - \beta_1[\mathrm{GenAI}]$ & 0.97 & 1.14 & 1.13 & 0.59 & [-0.04, 2.29] \\
\cline{2-8}

& LT $>$ AV & $\beta_1[\mathrm{LT}] - \beta_1[\mathrm{AV}]$ & 0.59 & 0.14 & 0.14 & 0.59 & [-1.03, 1.31] \\
\cline{2-8}

& AV $>$ GenAI & $\beta_1[\mathrm{AV}] - \beta_1[\mathrm{GenAI}]$ & 0.96 & 1.00 & 0.99 & 0.57 & [-0.17, 2.09] \\
\Xhline{2\arrayrulewidth}

\end{tabular}
}
\label{tab:diffse-quizscore-appendix}
\end{table*}

\begin{table*}[!ht]
\centering
\caption{Correlations}

\renewcommand{\arraystretch}{1.2}

\resizebox{\textwidth}{!}{%
\begin{tabular}{|c|c|c|c|c|c|c|}
\Xhline{2\arrayrulewidth}

\textbf{Measurement} & \textbf{Condition} & \textbf{Parameter} & \textbf{Mean} & \textbf{Median} & $\boldsymbol{\sigma}$ & \textbf{HDI} \\
\Xhline{2\arrayrulewidth}

\multirow{3}{*}{\centering Change in SE vs Quiz Score}
& LT & $\beta_1[\mathrm{LT}]$ & -0.11 & -0.10 & 0.11 & [-0.321, 0.113] \\
\cline{2-7}
& GenAI & $\beta_1[\mathrm{GenAI}]$ & -0.12 & -0.12 & 0.07 & [-0.259, 0.039] \\
\cline{2-7}
& AV & $\beta_1[\mathrm{AV}]$ & 0.21 & 0.21 & 0.10 & [0.015, 0.409] \\
\Xhline{2\arrayrulewidth}

\multirow{3}{*}{\centering \# Interactions vs Change in SE}
& LT & $\beta_1[\mathrm{LT}]$ & -0.59 & -0.59 & 0.44 & [-1.453, 0.277] \\
\cline{2-7}
& GenAI & $\beta_1[\mathrm{GenAI}]$ & -2.54 & -2.54 & 0.81 & [-4.090, -0.860] \\
\cline{2-7}
& AV & $\beta_1[\mathrm{AV}]$ & -0.60 & -0.60 & 0.59 & [-1.752, 0.555] \\
\Xhline{2\arrayrulewidth}

\multirow{3}{*}{\centering Interaction Time vs Change in SE}
& LT & $\beta_1[\mathrm{LT}]$ & -0.01 & -0.01 & 0.01 & [-0.030, 0.013] \\
\cline{2-7}
& GenAI & $\beta_1[\mathrm{GenAI}]$ & -0.04 & -0.04 & 0.02 & [-0.071, -0.001] \\
\cline{2-7}
& AV & $\beta_1[\mathrm{AV}]$ & -0.01 & -0.01 & 0.01 & [-0.031, 0.009] \\
\Xhline{2\arrayrulewidth}

\multirow{3}{*}{\centering \# Interactions vs Quiz Score}
& LT & $\beta_1[\mathrm{LT}]$ & -0.09 & -0.09 & 0.11 & [-0.300, 0.122] \\
\cline{2-7}
& GenAI & $\beta_1[\mathrm{GenAI}]$ & 0.21 & 0.21 & 0.23 & [-0.248, 0.656] \\
\cline{2-7}
& AV & $\beta_1[\mathrm{AV}]$ & 0.004 & 0.004 & 0.17 & [-0.326, 0.359] \\
\Xhline{2\arrayrulewidth}

\multirow{3}{*}{\centering Interaction Time vs Quiz Score}
& LT & $\beta_1[\mathrm{LT}]$ & -0.004 & -0.004 & 0.003 & [-0.009, 0.001] \\
\cline{2-7}
& GenAI & $\beta_1[\mathrm{GenAI}]$ & 0.005 & 0.005 & 0.005 & [-0.004, 0.014] \\
\cline{2-7}
& AV & $\beta_1[\mathrm{AV}]$ & -0.0003 & -0.0003 & 0.003 & [-0.006, 0.006] \\
\Xhline{2\arrayrulewidth}

\end{tabular}%
}

\label{tab:correlation-appendix}
\end{table*}

\newpage

\begin{table*}[!ht]
\centering
\caption{Post-session learning tool usefulness ratings}

\renewcommand{\arraystretch}{1.2}

\resizebox{\textwidth}{!}{%
\begin{tabular}{|c|c|c|c|c|c|c|c|}
\Xhline{2\arrayrulewidth}

\textbf{Measurement} & \textbf{Test} & \textbf{Parameter} & \textbf{Probability} & \textbf{Mean} & \textbf{Median} & $\boldsymbol{\sigma}$ & \textbf{HDI} \\
\Xhline{2\arrayrulewidth}

\multirow{6}{*}{\centering Designing}
& LT & $\beta_0[\mathrm{LT}] + \beta_1[\mathrm{LT}]$ & -- & 3.16 & 3.16 & 0.36 & [2.47, 3.86] \\
\cline{2-8}
& GenAI & $\beta_0[\mathrm{GenAI}] + \beta_1[\mathrm{GenAI}]$ & -- & 2.79 & 2.79 & 0.34 & [2.13, 3.47] \\
\cline{2-8}
& AV & $\beta_0[\mathrm{AV}] + \beta_1[\mathrm{AV}]$ & -- & 2.50 & 2.50 & 0.34 & [1.85, 3.18] \\
\cline{2-8}
& LT $>$ GenAI & $\beta_1[\mathrm{LT}]-\beta_1[\mathrm{GenAI}]$ & 0.78 & 0.37 & 0.37 & 0.49 & [-0.59, 1.32] \\
\cline{2-8}
& LT $>$ AV & $\beta_1[\mathrm{LT}]-\beta_1[\mathrm{AV}]$ & 0.92 & 0.66 & 0.66 & 0.49 & [-0.27, 1.66] \\
\cline{2-8}
& AV $>$ GenAI & $\beta_1[\mathrm{AV}]-\beta_1[\mathrm{GenAI}]$ & 0.26 & -0.29 & -0.28 & 0.47 & [-1.24, 0.62] \\
\Xhline{2\arrayrulewidth}

\multirow{6}{*}{\centering Understanding}
& LT & $\beta_0[\mathrm{LT}] + \beta_1[\mathrm{LT}]$ & -- & 3.47 & 3.47 & 0.30 & [2.89, 4.06] \\
\cline{2-8}
& GenAI & $\beta_0[\mathrm{GenAI}] + \beta_1[\mathrm{GenAI}]$ & -- & 2.96 & 2.96 & 0.27 & [2.42, 3.52] \\
\cline{2-8}
& AV & $\beta_0[\mathrm{AV}] + \beta_1[\mathrm{AV}]$ & -- & 2.58 & 2.58 & 0.27 & [2.02, 3.11] \\
\cline{2-8}
& LT $>$ GenAI & $\beta_1[\mathrm{LT}]-\beta_1[\mathrm{GenAI}]$ & 0.90 & 0.52 & 0.52 & 0.40 & [-0.28, 1.32] \\
\cline{2-8}
& LT $>$ AV & $\beta_1[\mathrm{LT}]-\beta_1[\mathrm{AV}]$ & 0.99 & 0.89 & 0.89 & 0.40 & [0.08, 1.68] \\
\cline{2-8}
& AV $>$ GenAI & $\beta_1[\mathrm{AV}]-\beta_1[\mathrm{GenAI}]$ & 0.16 & -0.37 & -0.37 & 0.39 & [-1.12, 0.41] \\
\Xhline{2\arrayrulewidth}

\multirow{6}{*}{\centering Running Time}
& LT & $\beta_0[\mathrm{LT}] + \beta_1[\mathrm{LT}]$ & -- & 3.14 & 3.14 & 0.42 & [2.34, 4.00] \\
\cline{2-8}
& GenAI & $\beta_0[\mathrm{GenAI}] + \beta_1[\mathrm{GenAI}]$ & -- & 3.00 & 3.00 & 0.39 & [2.22, 3.76] \\
\cline{2-8}
& AV & $\beta_0[\mathrm{AV}] + \beta_1[\mathrm{AV}]$ & -- & 1.75 & 1.75 & 0.39 & [0.96, 2.52] \\
\cline{2-8}
& LT $>$ GenAI & $\beta_1[\mathrm{LT}]-\beta_1[\mathrm{GenAI}]$ & 0.60 & 0.15 & 0.15 & 0.57 & [-0.97, 1.30] \\
\cline{2-8}
& LT $>$ AV & $\beta_1[\mathrm{LT}]-\beta_1[\mathrm{AV}]$ & 0.99 & 1.40 & 1.40 & 0.58 & [0.25, 2.50] \\
\cline{2-8}
& AV $>$ GenAI & $\beta_1[\mathrm{AV}]-\beta_1[\mathrm{GenAI}]$ & 0.01 & -1.25 & -1.25 & 0.55 & [-2.36, -0.18] \\
\Xhline{2\arrayrulewidth}

\multirow{6}{*}{\centering Implementation}
& LT & $\beta_0[\mathrm{LT}] + \beta_1[\mathrm{LT}]$ & -- & 3.23 & 3.23 & 0.39 & [2.46, 4.00] \\
\cline{2-8}
& GenAI & $\beta_0[\mathrm{GenAI}] + \beta_1[\mathrm{GenAI}]$ & -- & 3.09 & 3.09 & 0.37 & [2.36, 3.83] \\
\cline{2-8}
& AV & $\beta_0[\mathrm{AV}] + \beta_1[\mathrm{AV}]$ & -- & 2.75 & 2.75 & 0.37 & [2.03, 3.51] \\
\cline{2-8}
& LT $>$ GenAI & $\beta_1[\mathrm{LT}]-\beta_1[\mathrm{GenAI}]$ & 0.61 & 0.15 & 0.15 & 0.54 & [-0.94, 1.20] \\
\cline{2-8}
& LT $>$ AV & $\beta_1[\mathrm{LT}]-\beta_1[\mathrm{AV}]$ & 0.82 & 0.48 & 0.48 & 0.54 & [-0.59, 1.57] \\
\cline{2-8}
& AV $>$ GenAI & $\beta_1[\mathrm{AV}]-\beta_1[\mathrm{GenAI}]$ & 0.26 & -0.33 & -0.33 & 0.53 & [-1.38, 0.71] \\
\Xhline{2\arrayrulewidth}

\multirow{6}{*}{\centering Tracing Execution}
& LT & $\beta_0[\mathrm{LT}] + \beta_1[\mathrm{LT}]$ & -- & 3.51 & 3.51 & 0.41 & [2.73, 4.37] \\
\cline{2-8}
& GenAI & $\beta_0[\mathrm{GenAI}] + \beta_1[\mathrm{GenAI}]$ & -- & 3.13 & 3.13 & 0.39 & [2.37, 3.91] \\
\cline{2-8}
& AV & $\beta_0[\mathrm{AV}] + \beta_1[\mathrm{AV}]$ & -- & 3.58 & 3.58 & 0.39 & [2.80, 4.33] \\
\cline{2-8}
& LT $>$ GenAI & $\beta_1[\mathrm{LT}]-\beta_1[\mathrm{GenAI}]$ & 0.76 & 0.39 & 0.39 & 0.56 & [-0.692, 1.56] \\
\cline{2-8}
& LT $>$ AV & $\beta_1[\mathrm{LT}]-\beta_1[\mathrm{AV}]$ & 0.45 & -0.07 & -0.07 & 0.57 & [-1.18, 1.08] \\
\cline{2-8}
& AV $>$ GenAI & $\beta_1[\mathrm{AV}]-\beta_1[\mathrm{GenAI}]$ & 0.81 & 0.46 & 0.45 & 0.54 & [-0.65, 1.49] \\
\Xhline{2\arrayrulewidth}

\multirow{6}{*}{\centering Total}
& LT & $\beta_0[\mathrm{LT}] + \beta_1[\mathrm{LT}]$ & -- & 16.52 & 16.52 & 1.47 & [13.63, 19.45] \\
\cline{2-8}
& GenAI & $\beta_0[\mathrm{GenAI}] + \beta_1[\mathrm{GenAI}]$ & -- & 14.95 & 14.96 & 1.38 & [12.32, 17.80] \\
\cline{2-8}
& AV & $\beta_0[\mathrm{AV}] + \beta_1[\mathrm{AV}]$ & -- & 13.17 & 13.17 & 1.37 & [10.32, 15.75] \\
\cline{2-8}
& LT $>$ GenAI & $\beta_1[\mathrm{LT}]-\beta_1[\mathrm{GenAI}]$ & 0.79 & 1.56 & 1.55 & 2.02 & [-2.56, 5.44] \\
\cline{2-8}
& LT $>$ AV & $\beta_1[\mathrm{LT}]-\beta_1[\mathrm{AV}]$ & 0.95 & 3.36 & 3.35 & 2.02 & [-0.48, 7.41] \\
\cline{2-8}
& AV $>$ GenAI & $\beta_1[\mathrm{AV}]-\beta_1[\mathrm{GenAI}]$ & 0.17 & -1.79 & -1.78 & 1.94 & [-5.80, 1.91] \\
\Xhline{2\arrayrulewidth}

\end{tabular}%
}

\label{tab:usefulness_appendix}
\end{table*}

\newpage
\begin{table*}[!ht]
\centering
\caption{Final learning support comparative preference rankings}

\renewcommand{\arraystretch}{1.2}

\resizebox{\textwidth}{!}{%
\begin{tabular}{|c|c|c|c|c|c|c|c|}
\Xhline{2\arrayrulewidth}

\textbf{Measurement} & \textbf{Test} & \textbf{Parameter} & \textbf{Probability} & \textbf{Mean} & \textbf{Median} & $\boldsymbol{\sigma}$ & \textbf{HDI} \\
\Xhline{2\arrayrulewidth}

\multirow{6}{*}{\centering Designing}
& LT & $\beta_0[\mathrm{LT}] + \beta_1[\mathrm{LT}]$ & -- & 1.73 & 1.73 & 0.23 & [1.24, 2.16] \\
\cline{2-8}
& GenAI & $\beta_0[\mathrm{GenAI}] + \beta_1[\mathrm{GenAI}]$ & -- & 1.00 & 1.00 & 0.23 & [0.55, 1.48] \\
\cline{2-8}
& AV & $\beta_0[\mathrm{AV}] + \beta_1[\mathrm{AV}]$ & -- & 0.27 & 0.27 & 0.23 & [-0.19, 0.72] \\
\cline{2-8}
& LT $>$ GenAI & $\beta_1[\mathrm{LT}]-\beta_1[\mathrm{GenAI}]$ & 0.98 & 0.73 & 0.73 & 0.33 & [0.07, 1.38] \\
\cline{2-8}
& LT $>$ AV & $\beta_1[\mathrm{LT}]-\beta_1[\mathrm{AV}]$ & 0.99 & 1.45 & 1.46 & 0.33 & [0.77, 2.07] \\
\cline{2-8}
& AV $>$ GenAI & $\beta_1[\mathrm{AV}]-\beta_1[\mathrm{GenAI}]$ & 0.02 & -0.73 & -0.73 & 0.33 & [-1.35, -0.05] \\
\Xhline{2\arrayrulewidth}

\multirow{6}{*}{\centering Understanding}
& LT & $\beta_0[\mathrm{LT}] + \beta_1[\mathrm{LT}]$ & -- & 1.72 & 1.72 & 0.26 & [1.20, 2.24] \\
\cline{2-8}
& GenAI & $\beta_0[\mathrm{GenAI}] + \beta_1[\mathrm{GenAI}]$ & -- & 0.64 & 0.64 & 0.27 & [0.13, 1.18] \\
\cline{2-8}
& AV & $\beta_0[\mathrm{AV}] + \beta_1[\mathrm{AV}]$ & -- & 0.64 & 0.64 & 0.26 & [0.13, 1.17] \\
\cline{2-8}
& LT $>$ GenAI & $\beta_1[\mathrm{LT}]-\beta_1[\mathrm{GenAI}]$ & 0.99 & 1.08 & 1.08 & 0.37 & [0.36, 1.83] \\
\cline{2-8}
& LT $>$ AV & $\beta_1[\mathrm{LT}]-\beta_1[\mathrm{AV}]$ & 0.99 & 1.08 & 1.08 & 0.37 & [0.32, 1.78] \\
\cline{2-8}
& AV $>$ GenAI & $\beta_1[\mathrm{AV}]-\beta_1[\mathrm{GenAI}]$ & 0.50 & 0.00 & 0.00 & 0.37 & [-0.73, 0.74] \\
\Xhline{2\arrayrulewidth}

\multirow{6}{*}{\centering Running Time}
& LT & $\beta_0[\mathrm{LT}] + \beta_1[\mathrm{LT}]$ & -- & 1.27 & 1.28 & 0.28 & [0.72, 1.85] \\
\cline{2-8}
& GenAI & $\beta_0[\mathrm{GenAI}] + \beta_1[\mathrm{GenAI}]$ & -- & 1.36 & 1.36 & 0.28 & [0.80, 1.94] \\
\cline{2-8}
& AV & $\beta_0[\mathrm{AV}] + \beta_1[\mathrm{AV}]$ & -- & 0.37 & 0.37 & 0.28 & [-0.16, 0.93] \\
\cline{2-8}
& LT $>$ GenAI & $\beta_1[\mathrm{LT}]-\beta_1[\mathrm{GenAI}]$ & 0.41 & -0.09 & -0.09 & 0.40 & [-0.90, 0.68] \\
\cline{2-8}
& LT $>$ AV & $\beta_1[\mathrm{LT}]-\beta_1[\mathrm{AV}]$ & 0.99 & 0.91 & 0.91 & 0.40 & [0.08, 1.65] \\
\cline{2-8}
& AV $>$ GenAI & $\beta_1[\mathrm{AV}]-\beta_1[\mathrm{GenAI}]$ & 0.01 & -1.00 & -1.00 & 0.40 & [-1.74, -0.19] \\
\Xhline{2\arrayrulewidth}

\multirow{6}{*}{\centering Implementation}
& LT & $\beta_0[\mathrm{LT}] + \beta_1[\mathrm{LT}]$ & -- & 1.45 & 1.45 & 0.24 & [0.98, 1.92] \\
\cline{2-8}
& GenAI & $\beta_0[\mathrm{GenAI}] + \beta_1[\mathrm{GenAI}]$ & -- & 1.36 & 1.36 & 0.24 & [0.90, 1.84] \\
\cline{2-8}
& AV & $\beta_0[\mathrm{AV}] + \beta_1[\mathrm{AV}]$ & -- & 0.19 & 0.19 & 0.24 & [-0.29, 0.65] \\
\cline{2-8}
& LT $>$ GenAI & $\beta_1[\mathrm{LT}]-\beta_1[\mathrm{GenAI}]$ & 0.61 & 0.09 & 0.09 & 0.34 & [-0.57, 0.75] \\
\cline{2-8}
& LT $>$ AV & $\beta_1[\mathrm{LT}]-\beta_1[\mathrm{AV}]$ & 0.999 & 1.27 & 1.27 & 0.34 & [0.62, 1.98] \\
\cline{2-8}
& AV $>$ GenAI & $\beta_1[\mathrm{AV}]-\beta_1[\mathrm{GenAI}]$ & 0.001 & -1.17 & -1.18 & 0.34 & [-1.85, -0.52] \\
\Xhline{2\arrayrulewidth}

\multirow{6}{*}{\centering Tracing Execution}
& LT & $\beta_0[\mathrm{LT}] + \beta_1[\mathrm{LT}]$ & -- & 0.82 & 0.83 & 0.30 & [0.28, 1.43] \\
\cline{2-8}
& GenAI & $\beta_0[\mathrm{GenAI}] + \beta_1[\mathrm{GenAI}]$ & -- & 0.64 & 0.63 & 0.30 & [0.07, 1.23] \\
\cline{2-8}
& AV & $\beta_0[\mathrm{AV}] + \beta_1[\mathrm{AV}]$ & -- & 1.55 & 1.55 & 0.30 & [0.97, 2.14] \\
\cline{2-8}
& LT $>$ GenAI & $\beta_1[\mathrm{LT}]-\beta_1[\mathrm{GenAI}]$ & 0.68 & 0.19 & 0.19 & 0.42 & [-0.62, 1.04] \\
\cline{2-8}
& LT $>$ AV & $\beta_1[\mathrm{LT}]-\beta_1[\mathrm{AV}]$ & 0.05 & -0.72 & -0.73 & 0.42 & [-1.55, 0.12] \\
\cline{2-8}
& AV $>$ GenAI & $\beta_1[\mathrm{AV}]-\beta_1[\mathrm{GenAI}]$ & 0.98 & 0.91 & 0.91 & 0.42 & [0.09, 1.76] \\
\Xhline{2\arrayrulewidth}

\multirow{6}{*}{\centering Total}
& LT & $\beta_0[\mathrm{LT}] + \beta_1[\mathrm{LT}]$ & -- & 6.99 & 7.00 & 0.82 & [5.34, 8.56] \\
\cline{2-8}
& GenAI & $\beta_0[\mathrm{GenAI}] + \beta_1[\mathrm{GenAI}]$ & -- & 5.00 & 5.00 & 0.81 & [3.43, 6.62] \\
\cline{2-8}
& AV & $\beta_0[\mathrm{AV}] + \beta_1[\mathrm{AV}]$ & -- & 3.00 & 2.99 & 0.80 & [1.47, 4.66] \\
\cline{2-8}
& LT $>$ GenAI & $\beta_1[\mathrm{LT}]-\beta_1[\mathrm{GenAI}]$ & 0.96 & 1.99 & 2.00 & 1.16 & [-0.30, 4.18] \\
\cline{2-8}
& LT $>$ AV & $\beta_1[\mathrm{LT}]-\beta_1[\mathrm{AV}]$ & 0.999 & 3.99 & 4.00 & 1.14 & [1.77, 6.28] \\
\cline{2-8}
& AV $>$ GenAI & $\beta_1[\mathrm{AV}]-\beta_1[\mathrm{GenAI}]$ & 0.04 & -2.00 & -2.01 & 1.15 & [-4.33, 0.19] \\
\Xhline{2\arrayrulewidth}

\end{tabular}%
}

\label{tab:final-appendix}
\end{table*}

\end{document}